\documentclass[reprint,
superscriptaddress,
 amsmath,amssymb,
 aps,
pra,
]{revtex4-2}

\usepackage{graphicx}% Include figure files
\usepackage{dcolumn}% Align table columns on decimal point
\usepackage{xcolor}
\usepackage{physics}
\usepackage{bm}% bold math
\usepackage{hyperref}
\hypersetup{
	colorlinks = true,
	urlcolor   = blue,
	linkcolor  = blue,
	citecolor  = red
}

\begin{document}

\title{Multiparameter estimation of continuous-time Quantum Walk Hamiltonians through Machine Learning}% Force line breaks with \\

\author{Ilaria Gianani}
\affiliation{Dipartimento di Scienze, Universit\`{a} degli Studi Roma Tre, Via della Vasca Navale 84, 00146, Rome, Italy}

\author{Claudia Benedetti}
\affiliation{Dipartimento di Fisica ``Aldo Pontremoli'', Universit\`a degli Studi di Milano, 20133, Milan, Italy}

\begin{abstract}
The characterization of the Hamiltonian parameters defining a quantum walk is of paramount importance when performing a variety of tasks, from quantum communication to computation. When dealing with physical implementations of quantum walks, the parameters themselves may not be directly accessible, thus it is necessary to find alternative estimation strategies exploiting other observables. Here, we perform the multiparameter estimation of the Hamiltonian parameters characterizing a continuous-time quantum walk over a line graph with $n$-neighbour interactions using a deep neural network model fed with experimental probabilities at a given evolution time. We compare our results with the bounds derived from estimation theory and find that the neural network acts as a nearly optimal estimator both when the estimation of two or three parameters is performed.
%1) qw are an important tool per fare cose tipo computation e altri task\\
%2) however controlling the exact value of the H parameters from an experimental implementation can be hard with the required precision, thus one should perform a characterization. \\
%3) Here we show that we can successfully use a NN model per fare optimal estimation of H parameters of CTWQ with a fixed network defined by a varying number of parameters.\\
%4) our results are a first step towards achieving full control of experimental QWs qualcosa 
\end{abstract}

\maketitle

\section*{Introduction}

The characterization and control of quantum processes is a fundamental requirement for the realization of quantum protocols and crucial for the development of quantum technologies \cite{Eisert2020,Koch2022,Acin2018}. Controlling and validating the dynamics of quantum systems, i.e. their Hamiltonian, is a computationally-demanding task to perform with classical resources, but can be mitigated by the use of Machine Learning (ML) \cite{dunjko18,torlai20,Vernuccio2022}. 
ML techniques have been proven to be of substantial aid when adapted to quantum states characterization \cite{carleo17,bohrdt21,flynn22,Gutierrez22, vega22,koutny22}, optimization of control strategies \cite{mavadia17,niu19,fallani22,huang22}, quantum state transport \cite{porotti19,Brown_2021}, as well as for parameter estimation and classification tasks \cite{lumino18,cimini2019,ming19,Valeri2020,cimini2021,Ban21,palmieri21,gianani22}. 
Concerning the characterization of quantum processes, Hamiltonian learning strategies have been extensively investigated in order to provide a reliable solution to this challenge \cite{granade12,wiebe14,wiebe14A,Cao20,che21,rattacaso22}. Moreover, it is pivotal that the required information is often not directly accessible,  and must be inferred starting from experimental quantities \cite{jianwei17,gentile21}. 

For parametrized Hamiltonians, this translates in establishing the values of significant Hamiltonian parameters starting from measured quantities. This task is akin to what pertains quantum metrology \cite{Giovannetti2004,Giovannetti2011,PARIS2009}, and can be cast in terms of multiparameter estimation theory \cite{albarelli20}. In the estimation framework, Neural Networks (NN) have demonstrated to achieve superior performance in terms of reliability with finite-size data set and robustness to noise compared to standard estimators \cite{cimini2019} and as such they appear as a primary candidate for successfully estimating parametrized Hamiltonians. 

In the class of parametrized Hamiltonians, an exeptionally diffused and useful example is that of Quantum Walks (QWs). QWs are a universal and versatile tool that can be harnessed to perform a plethora of tasks ranging from energy transport \cite{Plenio_2008,mulken2011,uchiyama18} to quantum algorithms \cite{childs04,gamble10,Chakraborty20,atia21,paris21,candeloro22,apers22}, quantum computation \cite{childs09,lovett10,childs13}, and quantum communication \cite{bose07}. 
 In particular, continuous-time quantum walks (CTQWs) are the quantum analog of classical random walks \cite{fahri98,kempe003,venegas2012,kadian21} that describe the continuous evolution of a quantum particle over a set of discrete positions. These positions can be pairwise linked in different ways, generating graphs with different topologies. Indeed, the edges of a graph identify all the possible paths that the quantum particle can walk through. The edges weights, together with the on-site energies, are the relevant parameters characterizing the CTQW Hamiltonian. 
 
 CTQWs are especially useful to model physical phenomena such as  quantum  transport of energy in quantum biological systems, quantum routing and quantum state transfer \cite{Mulken:2006,kay10,Chudzicki10,kendon2011,alvir2016,cavazzoni22,Maquine22}. They have been both realized and simulated experimentally on different platforms, such as photons \cite{qiang16,tang18,imany20}, trapped atoms and ions \cite{meier16,tamura20}, waveguide arrays \cite{perets08,poulios14,caruso16,benedetti21}, microwaves \cite{bohm15}, and nuclear magnetic resonance \cite{du03}.

In all these tasks, fine tuning of the Hamiltonian parameters is required in order to achieve reliable and satisfactory results. The control of experimental CTQWs usually relies on a set of directly accessible experimental parameters which indirectly dictate the values of the Hamiltonian ones. However, since the mapping between these two sets of parameters can be particularly involved, in order to reliably characterizing the Hamiltonian, a detailed calibration linking one set to the other may not be sufficient. 
Hamiltonian learning strategies must hence be developed, estimating the relevant Hamiltonian parameters starting from experimental observables.

Here we show how to infer the Hamiltonian parameters of a CTQW unfolding on a line graph with $n$-neighbour interactions, having access only to the probability distribution on the graph's sites at a known time $t$ and to the initial state of the system. By casting the problem in informational terms, we determine the most suitable measurement configuration and then perform the estimation with a Deep Neural Network model. Our result show that our model acts as a nearly optimal estimator, saturating the bounds established by estimation theory.

\section*{Results}
{\bf \noindent Multiparameter estimation of CTQW.} CTQWs describe the evolution of a quantum particle that coherently moves among a set of $N_s$ discrete positions $\{\ket{x}\}_{x=1}^{N_s}$, which constitute a basis for the CTQW Hilbert space. 
The Hamiltonian generating the quantum dynamics is expressed in terms of on-site energies $\epsilon_x$ and couplings $J_{xy}$ between sites $x$ and $y$. Here we focus on a CTQW on a line with zero on-site energies $\epsilon_x=0\,\forall x$ and $n$-neighbour uniform couplings, such that the Hamiltonian can be writtes as:
\begin{equation}
%H=\sum_x \epsilon_x \ketbra{x}{x}+\sum_{xy}J_{xy}\ketbra{x}{y}.
H=-\sum_{i=1}^n\sum_{x}^{N_s-i} J_{i}\Big(\ketbra{x+i}{x}+\ketbra{x}{x+i}\Big).
\label{hami}
\end{equation}

The couplings $J_{i}$  are positive real numbers. This model could be generalized by considering complex couplings, which lead to chiral QWs \cite{Zimboras13,lu16,frigerio21,Khalique21,frigerio22}.
 Given the Hamiltonian \eqref{hami} and an initial state of the walker $\ket{\psi_0}$, the evolution of the CTQW at time $t$ is generated by the operator $e^{-i H t}$ such that the probability of occupying site $x$ is expressed as:
 \begin{align}
p_x(t,J_1,\dots,J_n)=|\bra{x}e^{-i H t}\ket{\psi_0}|^2.
\label{probx}
 \end{align}
The squared norm operation in Eq. \eqref{probx} establishes a non-linear mapping between its Hilbert space and the probability distributions in position space; this non-linearity, together with a high sensitivity to the initial condition, is linked to a chaotic behavior of QWs that can be exploited to build secure   cryptographyc protocols \cite{yang15,abdellatif2019,ellatif2020}.    
%As a consequence, determining the parameters of the Hamiltonian, having only access to the QW spatial probability distribution at a given time is not an easy task.

\begin{figure*}
    \includegraphics[width=\textwidth]{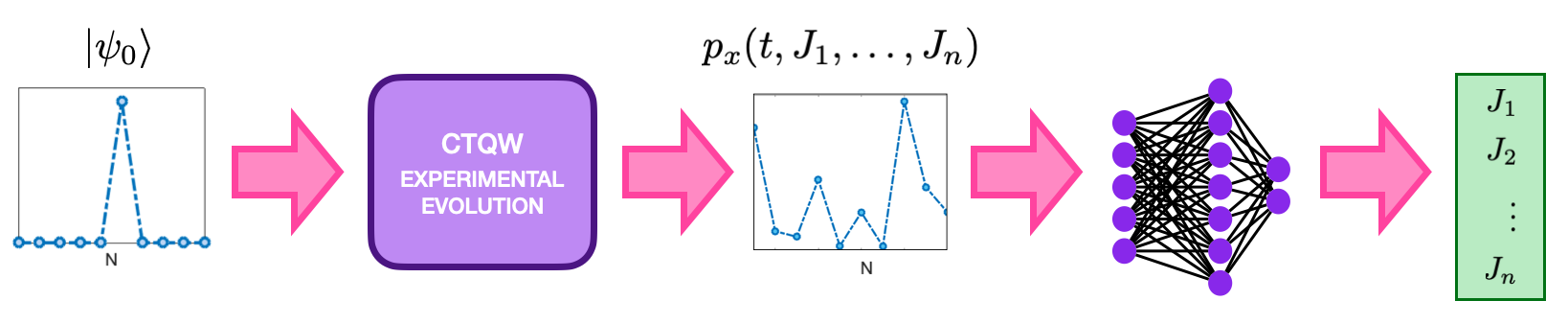}
    \caption{{\bf Conceptual scheme.} Starting from a given initial state, a system undergoes a CTQW and the position probabilities at a time $t$ are recorded. These are then used as input for a deep neural network which outputs the values of the Hamiltonian parameters defining the CTQW.}
    \label{fig:fig1}
\end{figure*}

The main objective of this work is described in Fig. \ref{fig:fig1}: our graph is a chain with $N_s$ sites, and the CTQW Hamiltonian is defined by $n$ parameters $J_i$, each varying in a known interval which we set as [0,1]. We assume to have control on the initial state $\vert \psi_0\rangle$ and to have access to the probability $p_x(t,J_1,\dots,J_{n})$ measured at a given time $t$. We aim at estimating the Hamiltonian parameters $J_1,...J_{n}$. As stated before, this problem is highly nonlinear, making a direct inversion a complex task. When the Hamiltonian depends only on one parameter, i.e. only uniform first-neighbour couplings are considered, it is possible to address the problem analytically \cite{tamascelli16,Seveso_2019}. Since we are dealing with an arbitrarily large but finite number of parameters we can cast this problem in terms of multiparemter estimation, and, given that our measurement strategy is fixed, {\it i.e.} we are performing a position measurement over all the sites of the chain, we can directly refer to the classical Fisher Information (FI). In this scenario, the FI is a matrix of dimension $n\times n$ whose elements are defined as: 

\begin{equation}
\label{eq:fisher}
  {F}_{ij}=\sum_x \frac{\partial_i p_x(t,J_1,\dots,J_{n}) \partial_j p_x(t,J_1,\dots,J_{n})}{p_x(t,J_1,\dots,J_{n})}. 
\end{equation}

By inverting the FI matrix, we can then cast the Carmér-Rao bound (CRB), lower bounding the variance of the estimated parameters $\Delta^2 J_i$ as: 

\begin{equation}
\label{eq:CRB}
M \cdot \Delta^2 J_{i} \geq {F}^{-1}_{ii},
\end{equation}
where M are the total resources employed for the measurement. This will serve as a reference to quantify the performance of our estimation. Estimation protocols conventionally make use of the  maximum likelihood or Bayesian strategies in order to derive an unbiased estimator \cite{PARIS2009,gianani20}. Machine Learning techniques have recently shown to provide a suitable alternative allowing to perform optimal estimation without recurring to a detailed model of the problem at hand. Specifically Neural Networks (NN) have been used both in single and multiparameter estimations demonstrating to successfully perform optimal estimation when trained with a sufficiently sampled dataset \cite{cimini2019,cimini2021}. Furthermore NN have proven higher robustness to noise compared to the other techniques. This black-box approach is particularly helpful in our scenario as, for most instances, $p_x(t,J_1,\dots,J_n)$ cannot be evaluated analytically. This complicates the use of traditional estimators in that the probabilities need to be sampled numerically. This issue is usually circumvented by employing Markov chain Monte Carlo methods \cite{spall03} to evaluate an estimator by sampling the numerical probability distributions, however contrary to the NN approach, these methods rely individually on punctual estimations of the probability distribution and may introduce biases in the estimation, which need to be accounted for \cite{Jacob2020}.

\begin{figure*}[b]
    \includegraphics[width=\textwidth]{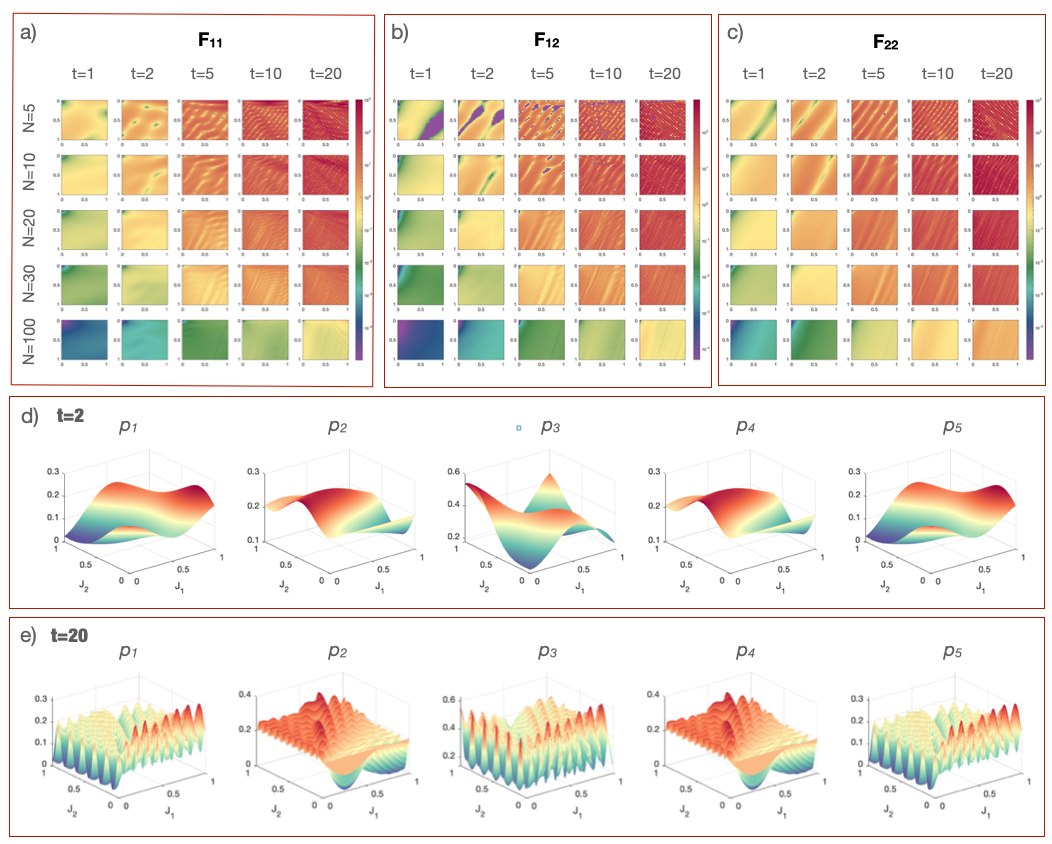}
    \caption{{\bf Information and probability space.} Fisher Information Matrix elements, $F_{11}$ (a), $F_{22}$(b), and $F_{33}$(c). Each 2D map is the FI element evaluated as function of the couplings $J_1$ and $J_2$ varying in the interval [0,1]. Each element is evaluated for different evolution times (columns) and chain lengths (rows), with $t=1,2,3,10,20$ and $N_s=5,10,20,30,100$. Panel (d) shows the five position probabilities $p_x$ for a 5-site chain as a function of the couplings $J_1$ and $J_2$ evaluated at time $t=2$, while panel (e) shows the same probabilities evaluated at time $t=20$. } 
    \label{fig:fisher}
\end{figure*}

The CRB in Eq. \ref{eq:CRB} will depend on all the Hamiltonian parameters $J_i$ as well on other parameters defining the evolution, i.e. the evolution time $t$, and the chain length $N_s$. 
Before moving on to the estimation of the Hamiltonian parameters $J_i$, we thus need to set a value for these other parameters by finding a viable regime in which to perform the estimation, dictated by the maximum amount of information extractable. For the sake of simplicity we consider $n=2$, so that only the first and second neighbour couplings can be active. This helps in the visualization of the results, but a similar analysis can be carried out with an arbitrary number of parameters. 
We consider five possible lengths for the chain, $N_s=5,10,20,30,100$ and five possible evolution times $t=1,2,5,10,20$ and evaluate the FI elements of Eq. \ref{eq:fisher} for each combination of $N_s$ and $t$. The derivatives in Eq. \ref{eq:fisher} are evaluated numerically. We consider as the initial state a 4th-order Gaussian with $\sigma=12$ centered at the center of the chain. This is fixed independently from the chain length. The results are shown in Fig. \ref{fig:fisher} in panels (a-c). Each plot in these panels is a 2D map of ${F}_{ij}$ as a function of $J_1$ and $J_2$. We can draw the following conclusions: at short times and for a short chain, the Fisher information is non zero for both couplings, and shows some structure depending on the parameters' values. At short times and for longer chains, the information decreases. This is because at longer lengths the initial state is not spread across the whole chain and we are performing a measurement over many sites which have not yet undergone any evolution and thus carry little information. This could be mitigated by employing different measurement strategies such as localized measurements addressing only the sites which have undergone the evolution \cite{tamascelli16}. We note that for longer chain lengths, the information on the second neighbour is consistently higher than that on the first neighbour. This is not surprising, because at any given time, the second neighbour information will have spanned a higher portion of the chain, and thus the information would travel faster compared to that of the first neighbour, hence mitigating the decrease in information previously discussed.
Finally, we note that increasing the evolution time corresponds to increasing the amount of information, at the cost of the information becoming heavily structured. This reflects the shape of the probability space, shown in Fig. \ref{fig:fisher} panel (d-e), at $t=2$ and $t=20$ for $N_s=5$. \\

{\bf \noindent Two-parameter estimation.} We now proceed with the estimation, while keeping n=2. Based on the results of Fig. \ref{fig:fisher}, we now seek for a combination of chain length and evolution time where the information on the sought parameters is sufficiently high but not overly structured. Indeed, while it is desirable to have a high information content, a heavily structured profile would be detrimental to the estimation, requiring a tighter sampling of the NN's training dataset. We hence fix our chain length to $N_s=10$ sites and set the evolution time at t=1.5. 

In order to perform the estimation we implement a deep neural network model as follows: the input features are the probabilities of detection at each site $x$ of the chain $p_x(t,J_1,J_2)$, hence the input layer is comprised of $N_s=10$ neurons. We then normalize the input features using a Batch Normalization layer which is followed by six hidden layers of 600 neurons each. The network outputs the value of the two couplings, hence the output layer will consist of two neurons. 

\begin{figure*}
    \includegraphics[width=\textwidth]{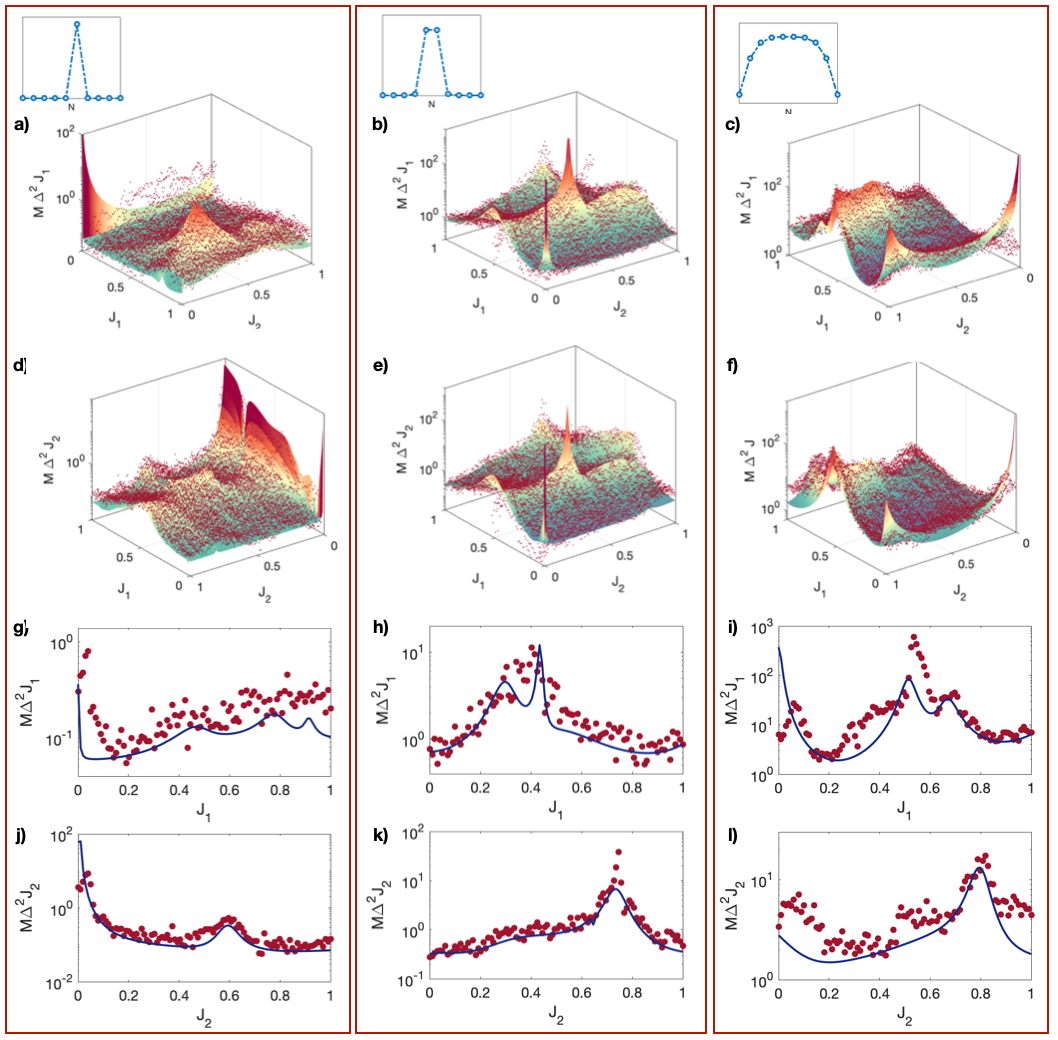}
    \caption{{\bf Two-parameter estimation.} Results of the estimation of the first and second neighbour couplings performed using the NN model with three different input states, shown in the inset (columns). Panels (a-c) show the log-plots of CRB for $J_1$ (surface) and the variance of the NN prediction evaluated over the Monte Carlo samples multiplied by the total number of resources M (red circles - see text). Panels (d-f) show the log-plots of CRB (surface) and evaluated variance (red circles) for $J_2$. Panels (g-i) show a slice of the 3D plot of $M \Delta^2 J_1$ for $J_2=1$, while panels (j-l) show a slice of the 3D plot of $M \Delta^2J_2$ for $J_1=1$. In these the blue line is the CRB and the red dots are the estimated variances. }
    \label{fig:CRB2}
\end{figure*}

We consider three different initial states $\vert\psi_0\rangle$: a Gaussian with $\sigma= 0.2$ centered on site $x=5$ corresponding to a localized excitation, a Gaussian with $\sigma= 0.5$ centered at the center of the chain corresponding to an excitation mostly divided between two sites of the graph, and a 4th-order Gaussian with $\sigma=12$ centered at the center of the chain corresponding to an excitation spread across the whole graph. These are shown in the insets of Fig. \ref{fig:CRB2}. Once we have selected the input state we perform the following procedure for training and testing:

\begin{itemize}
\item[1.] we generate $N_{samp}=2^{14}$ random values for the couple $\{J_1,J_2\}$ uniformly distributed in the extended interval $[-0.2,1.2]$ in order to limit the bias due to boundary conditions \cite{cimini2019};
\item[2.] we use the generated couplings to perform the evolution at time t=1.5 according to Eq.\ref{probx} and record the probability $p_x(t,J_1,J_2)$ for each site $x$ in the chain;
\item[3.] we generate simulated counts multiplying the probability $p_x(t,J_1,J_2)$ by the total number of resources which we set at $M=2\cdot10^5$, so that $\pi_x(t,J_1,J_2)= M \cdot p_x(t,J_1,J_2)$. This puts us in a regime where the CRB should be saturated;
\item[4.]  in order to account for fluctuations we bootstrap the training data by running $N_{MC}=500$ Monte Carlo routines extracting new values $\pi_x'(t,J_1,J_2)$ from a Poisson distribution of mean $\pi_x(t,J_1,J_2)$;
\item[5.] we split the generated dataset into the training (0.8) and validation set (0.2); 
\item[6.] we run the training for 200 epochs with a batch size of 1000 using the Adam optimizer with learning rate set to $10^{-3}$ and adopting as metric the MSE. 
\end{itemize}
In order to test the network we generate a new set of $N_{test}=10^4$ values of the couples $\{J_1,J_2\}$, now in $[0,1]$, and evaluate the evolved probabilities, which we multiply by $M$ to simulate the measured counts. We then perform 300 Monte Carlo runs to account for Poissonian noise and use the NN model to predict the values of the couplings for each generated probability. We can then calculate the error on the estimated couplings as the variance $\Delta^2 J_i$ over the Monte Carlo samples and compare it with the CRB, which we obtain from Eq. \ref{eq:CRB} by inverting the FI matrix of Eq. \ref{eq:fisher}. As for the analysis of the previous section, the derivatives of Eq. \ref{eq:fisher} are evaluated numerically. 

 The results of the estimation are shown in Fig. \ref{fig:CRB2}. The plots show the agreement between the CRB (surfaces) with the variance on estimated values $\Delta^2 J_i$ multiplied by the number of resources M (red dots), indicating that the NN model is able to perform a nearly optimal estimation. The graphs in panels (g-l) are slices of the 3D plots for the errors on the first (second) parameter, when the second (first) parameter is taken to be equal to its maximum value, $J_{2(1)}= 1$. The NN model performs comparably for all input states considered.  \\

{\bf \noindent Three-parameter estimation.} We now extend the estimation to the case where the CTQW Hamiltonian is defined by three parameters, by considering also the third-neighbour coupling $J_3$. We keep the chain length fixed at $N_s=10$ and the evolution time at $t=1.5$. Since there was no significant discrepancy between the estimations starting with different $\vert \psi_0\rangle$, we only consider the localized state, i.e. the Gaussian with $\sigma=0.2$. The NN model we use is the same as for the two parameter estimation, the only difference being the number of neurons in the output layer, which now amounts to three. 
The training procedure too remains the same, but since the parameters space has increased, so must the size of the training dataset. We thus now consider $N_{samp}=2^{18}$ randomly generated triplets $\{J_1,J_2,J_3\}$, while keeping all the other hyperparameters unchanged.
As before, we numerically evaluate the CRB from Eq \ref{eq:CRB} starting from the Fisher information matrix of Eq. \ref{eq:fisher}.

\begin{figure*}
    \includegraphics[width=\textwidth]{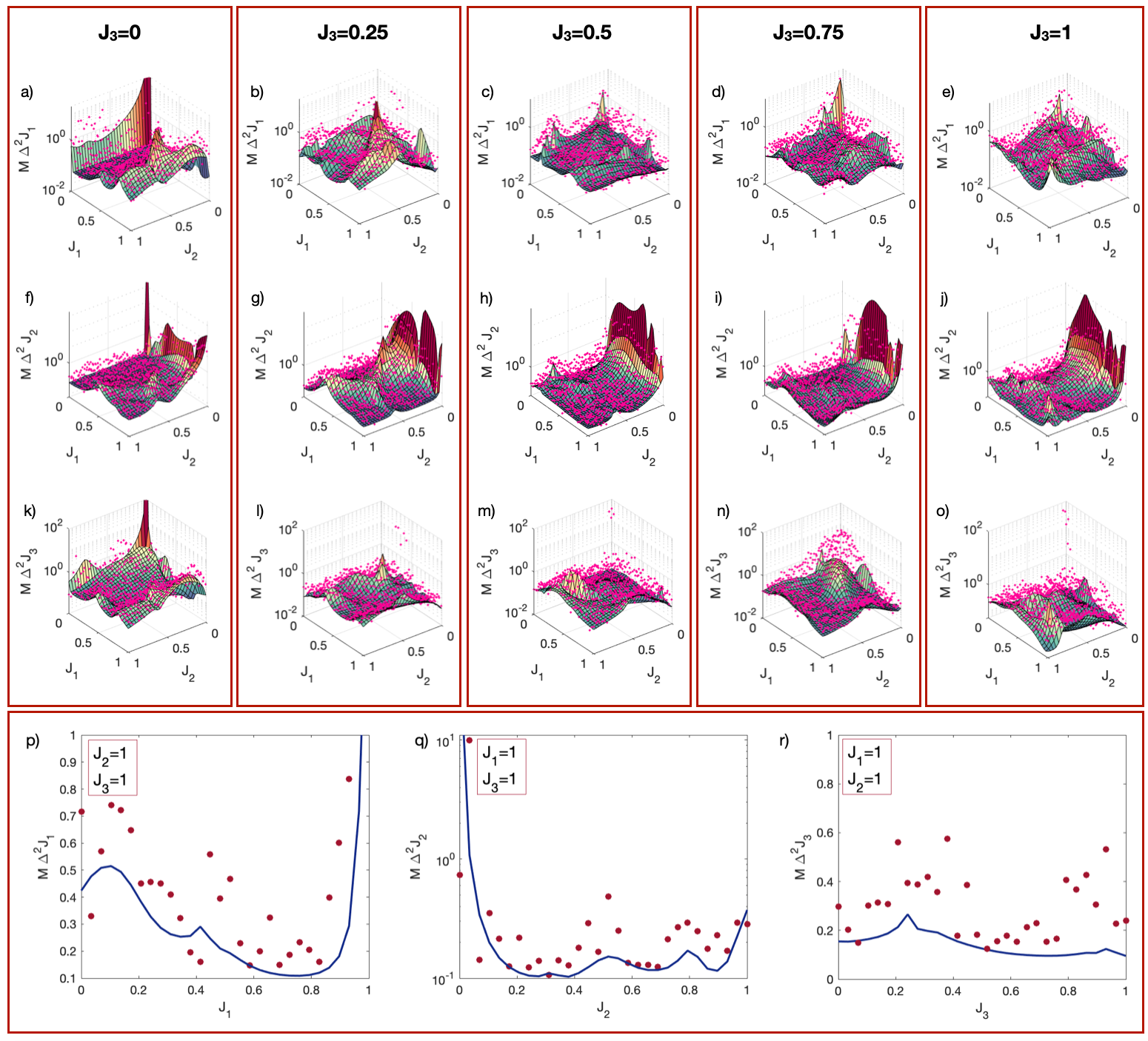}
    \caption{{\bf Three-parameter estimation.} Results of the estimation of the first, second and third-neighbour couplings using the NN model for a fixed input state. Each 3D log-plot is a map showing the CRB (surface) and estimated variance (fuchsia circles) for $M\Delta^2J_1$ (panels a-e),$M\Delta^2J_2$ (panels f-j), and $M\Delta^2J_3$ (panels k-o) as a function of $J_1$ and $J_2$ for different values of $J_3=0,0.25,0.5,0.75,1$ for each column. Panels (p-r) shows the CRB (blue line) and evaluated variances (red circles) for $M\Delta^2J_1$ as a function of $J_1$ for $J_2=J_3=1$ (panel p), $M\Delta^2J_2$ as a function of $J_2$ for $J_1=J_3=1$ (panel q), and $M\Delta^2J_3$ as a function of $J_3$ for $J_1=J_2=1$ (panel r).}
    \label{fig:CRB3}
\end{figure*}

The comparison between the CRB and the estimated values is shown in Fig. \ref{fig:CRB3}. The error on each $J_i$ will now depend on all three couplings $\{J_1,J_2,J_3\}$. In order to keep the consistency with the previous example, we report the error as a map of the first two couplings, for different values of the third one, one for each column of Fig. \ref{fig:CRB3}. As before, the surfaces indicate the CRB value and the dots (fuchsia) are the variance $\Delta^2 J_i$ on the estimated points multiplied by the number of resources M.
Panels (p-r) show slices of the graphs for the error on each parameter $J_i$ while the other two are kept at their maximum value, $J_{j,k}=1$. The plots still show a good agreement between the estimated values and the CRB, demonstrating a nearly optimal estimation in this instance as well. This comes however at the cost of substantially increasing the dimension of training set, which is expected as the parameter space increases. 

\section*{Discussion}

 Characterizing quantum processes with the utmost attainable precision is of paramount importance in order to harness them for various tasks. Indeed, errors in the calibration would reverberate in the task at hand, hindering the reliability of the final result, and as such they need to be validated and kept at a minimum.

 Here we have presented a multiparameter estimation-based approach for characterizing parametrized Hamiltonians, focusing on the dynamics of CTQWs. 
 Starting solely from experimentally measurable quantities, i.e. the position distribution across the graph's sites, our method allows to perform nearly-optimal estimation of the Hamiltonian parameters. We have demonstrated our approach by performing the estimation for a CTQW on a line graph for $n$-neighbour couplings with $n=2$ and $n=3$.
 
 The estimation is carried out using a deep neural network model which we train with data simulating experimental measured counts. These are then bootstrapped to account for Poissonian noise. Our method achieves a reliable estimation nearly saturating the CRB both for $n=2$ and $n=3$. 
 We have explored the informational content of the chosen measurement strategy as a function of the graph dimension and of the evolution time at which the measurement is performed. This highlights how the measurement strategy over all sites of the graph performs better when all sites have already undergone an evolution at a fixed time. If this is not the case, localized measurements should be investigated. This also shows how, while longer times provide a higher information content, the highly structured profile of the probabilities (and hence of the FI) would impose a tighter sampling of the training set for the NN to successfully reconstruct the profile. 
 From panels (g-l) of Fig. \ref{fig:CRB2} we note how, while the estimated points follow closely the less-featured values of the CRB, the estimation does slightly deteriorates when there are two close features. As shown before \cite{cimini2019} this is also critically related to the sampling rate of the training set, which can be adjusted to attain the desired accuracy. It is possible to further increase the number of parameters by adapting the size of the training data set. This does affect the computational resources, but can be easily managed adopting well-established techniques such as batch generators. This may allow also to extend these results to multiple particles and explore more complex topologies.\\

\section*{Aknowledgements}

We acknowledge M. Barbieri, M. Genoni, F. Albarelli and M. Paris for fruitful discussion. I.G. acknowledges the support of the project FET-OPEN-RIA STORMYTUNE (Grant Agreement No. 899587).
CB acknowledges support from the Sviluppo UniMi 2021 initiative.
\section*{Author Contributions}
I.G. and C.B. conceived the project, I.G. carried out the analysis. Both authors discussed the results and contributed to writing the manuscript.

\bibliography{mainbib}% Produces the bibliography via BibTeX.

\begin{thebibliography}{87}%
\makeatletter
\providecommand \@ifxundefined [1]{%
 \@ifx{#1\undefined}
}%
\providecommand \@ifnum [1]{%
 \ifnum #1\expandafter \@firstoftwo
 \else \expandafter \@secondoftwo
 \fi
}%
\providecommand \@ifx [1]{%
 \ifx #1\expandafter \@firstoftwo
 \else \expandafter \@secondoftwo
 \fi
}%
\providecommand \natexlab [1]{#1}%
\providecommand \enquote  [1]{``#1''}%
\providecommand \bibnamefont  [1]{#1}%
\providecommand \bibfnamefont [1]{#1}%
\providecommand \citenamefont [1]{#1}%
\providecommand \href@noop [0]{\@secondoftwo}%
\providecommand \href [0]{\begingroup \@sanitize@url \@href}%
\providecommand \@href[1]{\@@startlink{#1}\@@href}%
\providecommand \@@href[1]{\endgroup#1\@@endlink}%
\providecommand \@sanitize@url [0]{\catcode `\\12\catcode `\$12\catcode
  `\&12\catcode `\#12\catcode `\^12\catcode `\_12\catcode `\%12\relax}%
\providecommand \@@startlink[1]{}%
\providecommand \@@endlink[0]{}%
\providecommand \url  [0]{\begingroup\@sanitize@url \@url }%
\providecommand \@url [1]{\endgroup\@href {#1}{\urlprefix }}%
\providecommand \urlprefix  [0]{URL }%
\providecommand \Eprint [0]{\href }%
\providecommand \doibase [0]{https://doi.org/}%
\providecommand \selectlanguage [0]{\@gobble}%
\providecommand \bibinfo  [0]{\@secondoftwo}%
\providecommand \bibfield  [0]{\@secondoftwo}%
\providecommand \translation [1]{[#1]}%
\providecommand \BibitemOpen [0]{}%
\providecommand \bibitemStop [0]{}%
\providecommand \bibitemNoStop [0]{.\EOS\space}%
\providecommand \EOS [0]{\spacefactor3000\relax}%
\providecommand \BibitemShut  [1]{\csname bibitem#1\endcsname}%
\let\auto@bib@innerbib\@empty
%</preamble>
\bibitem [{\citenamefont {Eisert}\ \emph {et~al.}(2020)\citenamefont {Eisert},
  \citenamefont {Hangleiter}, \citenamefont {Walk}, \citenamefont {Roth},
  \citenamefont {Markham}, \citenamefont {Parekh}, \citenamefont {Chabaud},\
  and\ \citenamefont {Kashefi}}]{Eisert2020}%
  \BibitemOpen
  \bibfield  {author} {\bibinfo {author} {\bibfnamefont {J.}~\bibnamefont
  {Eisert}}, \bibinfo {author} {\bibfnamefont {D.}~\bibnamefont {Hangleiter}},
  \bibinfo {author} {\bibfnamefont {N.}~\bibnamefont {Walk}}, \bibinfo {author}
  {\bibfnamefont {I.}~\bibnamefont {Roth}}, \bibinfo {author} {\bibfnamefont
  {D.}~\bibnamefont {Markham}}, \bibinfo {author} {\bibfnamefont
  {R.}~\bibnamefont {Parekh}}, \bibinfo {author} {\bibfnamefont
  {U.}~\bibnamefont {Chabaud}},\ and\ \bibinfo {author} {\bibfnamefont
  {E.}~\bibnamefont {Kashefi}},\ }\bibfield  {title} {\bibinfo {title}
  {{Quantum certification and benchmarking}},\ }\href
  {https://doi.org/10.1038/s42254-020-0186-4} {\bibfield  {journal} {\bibinfo
  {journal} {Nat. Rev. Phys.}\ }\textbf {\bibinfo {volume} {2}},\ \bibinfo
  {pages} {382} (\bibinfo {year} {2020})}\BibitemShut {NoStop}%
\bibitem [{\citenamefont {Koch}\ \emph {et~al.}(2022)\citenamefont {Koch},
  \citenamefont {Boscain}, \citenamefont {Calarco}, \citenamefont {Dirr},
  \citenamefont {Filipp}, \citenamefont {Glaser}, \citenamefont {Kosloff},
  \citenamefont {Montangero}, \citenamefont {Schulte-Herbr{\"{u}}ggen},
  \citenamefont {Sugny},\ and\ \citenamefont {Wilhelm}}]{Koch2022}%
  \BibitemOpen
  \bibfield  {author} {\bibinfo {author} {\bibfnamefont {C.~P.}\ \bibnamefont
  {Koch}}, \bibinfo {author} {\bibfnamefont {U.}~\bibnamefont {Boscain}},
  \bibinfo {author} {\bibfnamefont {T.}~\bibnamefont {Calarco}}, \bibinfo
  {author} {\bibfnamefont {G.}~\bibnamefont {Dirr}}, \bibinfo {author}
  {\bibfnamefont {S.}~\bibnamefont {Filipp}}, \bibinfo {author} {\bibfnamefont
  {S.~J.}\ \bibnamefont {Glaser}}, \bibinfo {author} {\bibfnamefont
  {R.}~\bibnamefont {Kosloff}}, \bibinfo {author} {\bibfnamefont
  {S.}~\bibnamefont {Montangero}}, \bibinfo {author} {\bibfnamefont
  {T.}~\bibnamefont {Schulte-Herbr{\"{u}}ggen}}, \bibinfo {author}
  {\bibfnamefont {D.}~\bibnamefont {Sugny}},\ and\ \bibinfo {author}
  {\bibfnamefont {F.~K.}\ \bibnamefont {Wilhelm}},\ }\bibfield  {title}
  {\bibinfo {title} {{Quantum optimal control in quantum technologies.
  Strategic report on current status, visions and goals for research in
  Europe}},\ }\href {https://doi.org/10.1140/epjqt/s40507-022-00138-x}
  {\bibfield  {journal} {\bibinfo  {journal} {EPJ Quantum Technol.}\ }\textbf
  {\bibinfo {volume} {9}},\ \bibinfo {pages} {19} (\bibinfo {year}
  {2022})}\BibitemShut {NoStop}%
\bibitem [{\citenamefont {Acín}\ \emph {et~al.}(2018)\citenamefont {Acín},
  \citenamefont {Bloch}, \citenamefont {Buhrman}, \citenamefont {Calarco},
  \citenamefont {Eichler}, \citenamefont {Eisert}, \citenamefont {Esteve},
  \citenamefont {Gisin}, \citenamefont {Glaser}, \citenamefont {Jelezko},
  \citenamefont {Kuhr}, \citenamefont {Lewenstein}, \citenamefont {Riedel},
  \citenamefont {Schmidt}, \citenamefont {Thew}, \citenamefont {Wallraff},
  \citenamefont {Walmsley},\ and\ \citenamefont {Wilhelm}}]{Acin2018}%
  \BibitemOpen
  \bibfield  {author} {\bibinfo {author} {\bibfnamefont {A.}~\bibnamefont
  {Acín}}, \bibinfo {author} {\bibfnamefont {I.}~\bibnamefont {Bloch}},
  \bibinfo {author} {\bibfnamefont {H.}~\bibnamefont {Buhrman}}, \bibinfo
  {author} {\bibfnamefont {T.}~\bibnamefont {Calarco}}, \bibinfo {author}
  {\bibfnamefont {C.}~\bibnamefont {Eichler}}, \bibinfo {author} {\bibfnamefont
  {J.}~\bibnamefont {Eisert}}, \bibinfo {author} {\bibfnamefont
  {D.}~\bibnamefont {Esteve}}, \bibinfo {author} {\bibfnamefont
  {N.}~\bibnamefont {Gisin}}, \bibinfo {author} {\bibfnamefont {S.~J.}\
  \bibnamefont {Glaser}}, \bibinfo {author} {\bibfnamefont {F.}~\bibnamefont
  {Jelezko}}, \bibinfo {author} {\bibfnamefont {S.}~\bibnamefont {Kuhr}},
  \bibinfo {author} {\bibfnamefont {M.}~\bibnamefont {Lewenstein}}, \bibinfo
  {author} {\bibfnamefont {M.~F.}\ \bibnamefont {Riedel}}, \bibinfo {author}
  {\bibfnamefont {P.~O.}\ \bibnamefont {Schmidt}}, \bibinfo {author}
  {\bibfnamefont {R.}~\bibnamefont {Thew}}, \bibinfo {author} {\bibfnamefont
  {A.}~\bibnamefont {Wallraff}}, \bibinfo {author} {\bibfnamefont
  {I.}~\bibnamefont {Walmsley}},\ and\ \bibinfo {author} {\bibfnamefont
  {F.~K.}\ \bibnamefont {Wilhelm}},\ }\bibfield  {title} {\bibinfo {title} {The
  quantum technologies roadmap: a european community view},\ }\href
  {https://doi.org/10.1088/1367-2630/aad1ea} {\bibfield  {journal} {\bibinfo
  {journal} {New J. Phys.}\ }\textbf {\bibinfo {volume} {20}},\ \bibinfo
  {pages} {080201} (\bibinfo {year} {2018})}\BibitemShut {NoStop}%
\bibitem [{\citenamefont {Dunjko}\ and\ \citenamefont
  {Briegel}(2018)}]{dunjko18}%
  \BibitemOpen
  \bibfield  {author} {\bibinfo {author} {\bibfnamefont {V.}~\bibnamefont
  {Dunjko}}\ and\ \bibinfo {author} {\bibfnamefont {H.~J.}\ \bibnamefont
  {Briegel}},\ }\bibfield  {title} {\bibinfo {title} {Machine learning and
  artificial intelligence in the quantum domain: a review of recent progress},\
  }\href {https://doi.org/10.1088/1361-6633/aab406} {\bibfield  {journal}
  {\bibinfo  {journal} {Rep. Prog. Phys.}\ }\textbf {\bibinfo {volume} {81}},\
  \bibinfo {pages} {074001} (\bibinfo {year} {2018})}\BibitemShut {NoStop}%
\bibitem [{\citenamefont {Torlai}\ and\ \citenamefont
  {Melko}(2020)}]{torlai20}%
  \BibitemOpen
  \bibfield  {author} {\bibinfo {author} {\bibfnamefont {G.}~\bibnamefont
  {Torlai}}\ and\ \bibinfo {author} {\bibfnamefont {R.~G.}\ \bibnamefont
  {Melko}},\ }\bibfield  {title} {\bibinfo {title} {Machine-learning quantum
  states in the nisq era},\ }\href
  {https://doi.org/10.1146/annurev-conmatphys-031119-050651} {\bibfield
  {journal} {\bibinfo  {journal} {Annu. Rev. Condens. Matter Phys.}\ }\textbf
  {\bibinfo {volume} {11}},\ \bibinfo {pages} {325} (\bibinfo {year}
  {2020})}\BibitemShut {NoStop}%
\bibitem [{\citenamefont {Vernuccio}\ \emph {et~al.}(2022)\citenamefont
  {Vernuccio}, \citenamefont {Bresci}, \citenamefont {Cimini}, \citenamefont
  {Giuseppi}, \citenamefont {Cerullo}, \citenamefont {Polli},\ and\
  \citenamefont {Valensise}}]{Vernuccio2022}%
  \BibitemOpen
  \bibfield  {author} {\bibinfo {author} {\bibfnamefont {F.}~\bibnamefont
  {Vernuccio}}, \bibinfo {author} {\bibfnamefont {A.}~\bibnamefont {Bresci}},
  \bibinfo {author} {\bibfnamefont {V.}~\bibnamefont {Cimini}}, \bibinfo
  {author} {\bibfnamefont {A.}~\bibnamefont {Giuseppi}}, \bibinfo {author}
  {\bibfnamefont {G.}~\bibnamefont {Cerullo}}, \bibinfo {author} {\bibfnamefont
  {D.}~\bibnamefont {Polli}},\ and\ \bibinfo {author} {\bibfnamefont {C.~M.}\
  \bibnamefont {Valensise}},\ }\bibfield  {title} {\bibinfo {title}
  {{Artificial Intelligence in Classical and Quantum Photonics}},\ }\href
  {https://doi.org/https://doi.org/10.1002/lpor.202100399} {\bibfield
  {journal} {\bibinfo  {journal} {Laser Photonics Rev.}\ }\textbf {\bibinfo
  {volume} {16}},\ \bibinfo {pages} {2100399} (\bibinfo {year}
  {2022})}\BibitemShut {NoStop}%
\bibitem [{\citenamefont {Carleo}\ and\ \citenamefont
  {Troyer}(2017)}]{carleo17}%
  \BibitemOpen
  \bibfield  {author} {\bibinfo {author} {\bibfnamefont {G.}~\bibnamefont
  {Carleo}}\ and\ \bibinfo {author} {\bibfnamefont {M.}~\bibnamefont
  {Troyer}},\ }\bibfield  {title} {\bibinfo {title} {Solving the quantum
  many-body problem with artificial neural networks},\ }\href
  {https://doi.org/10.1126/science.aag2302} {\bibfield  {journal} {\bibinfo
  {journal} {Science}\ }\textbf {\bibinfo {volume} {355}},\ \bibinfo {pages}
  {602} (\bibinfo {year} {2017})}\BibitemShut {NoStop}%
\bibitem [{\citenamefont {Bohrdt}\ \emph {et~al.}(2021)\citenamefont {Bohrdt},
  \citenamefont {Kim}, \citenamefont {Lukin}, \citenamefont {Rispoli},
  \citenamefont {Schittko}, \citenamefont {Knap}, \citenamefont {Greiner},\
  and\ \citenamefont {L\'eonard}}]{bohrdt21}%
  \BibitemOpen
  \bibfield  {author} {\bibinfo {author} {\bibfnamefont {A.}~\bibnamefont
  {Bohrdt}}, \bibinfo {author} {\bibfnamefont {S.}~\bibnamefont {Kim}},
  \bibinfo {author} {\bibfnamefont {A.}~\bibnamefont {Lukin}}, \bibinfo
  {author} {\bibfnamefont {M.}~\bibnamefont {Rispoli}}, \bibinfo {author}
  {\bibfnamefont {R.}~\bibnamefont {Schittko}}, \bibinfo {author}
  {\bibfnamefont {M.}~\bibnamefont {Knap}}, \bibinfo {author} {\bibfnamefont
  {M.}~\bibnamefont {Greiner}},\ and\ \bibinfo {author} {\bibfnamefont
  {J.}~\bibnamefont {L\'eonard}},\ }\bibfield  {title} {\bibinfo {title}
  {Analyzing nonequilibrium quantum states through snapshots with artificial
  neural networks},\ }\href {https://doi.org/10.1103/PhysRevLett.127.150504}
  {\bibfield  {journal} {\bibinfo  {journal} {Phys. Rev. Lett.}\ }\textbf
  {\bibinfo {volume} {127}},\ \bibinfo {pages} {150504} (\bibinfo {year}
  {2021})}\BibitemShut {NoStop}%
\bibitem [{\citenamefont {Flynn}\ \emph {et~al.}(2022)\citenamefont {Flynn},
  \citenamefont {Gentile}, \citenamefont {Wiebe}, \citenamefont {Santagati},\
  and\ \citenamefont {Laing}}]{flynn22}%
  \BibitemOpen
  \bibfield  {author} {\bibinfo {author} {\bibfnamefont {B.}~\bibnamefont
  {Flynn}}, \bibinfo {author} {\bibfnamefont {A.~A.}\ \bibnamefont {Gentile}},
  \bibinfo {author} {\bibfnamefont {N.}~\bibnamefont {Wiebe}}, \bibinfo
  {author} {\bibfnamefont {R.}~\bibnamefont {Santagati}},\ and\ \bibinfo
  {author} {\bibfnamefont {A.}~\bibnamefont {Laing}},\ }\bibfield  {title}
  {\bibinfo {title} {Quantum model learning agent: characterisation of quantum
  systems through machine learning},\ }\href
  {https://doi.org/10.1088/1367-2630/ac68ff} {\bibfield  {journal} {\bibinfo
  {journal} {New J. Phys.}\ }\textbf {\bibinfo {volume} {24}},\ \bibinfo
  {pages} {053034} (\bibinfo {year} {2022})}\BibitemShut {NoStop}%
\bibitem [{\citenamefont {Guti{\'{e}}rrez}\ and\ \citenamefont
  {Mendl}(2022)}]{Gutierrez22}%
  \BibitemOpen
  \bibfield  {author} {\bibinfo {author} {\bibfnamefont {I.~L.}\ \bibnamefont
  {Guti{\'{e}}rrez}}\ and\ \bibinfo {author} {\bibfnamefont {C.~B.}\
  \bibnamefont {Mendl}},\ }\bibfield  {title} {\bibinfo {title} {Real time
  evolution with neural-network quantum states},\ }\href
  {https://doi.org/10.22331/q-2022-01-20-627} {\bibfield  {journal} {\bibinfo
  {journal} {{Quantum}}\ }\textbf {\bibinfo {volume} {6}},\ \bibinfo {pages}
  {627} (\bibinfo {year} {2022})}\BibitemShut {NoStop}%
\bibitem [{\citenamefont {Papi\ifmmode~\check{c}\else \v{c}\fi{}}\ and\
  \citenamefont {de~Vega}(2022)}]{vega22}%
  \BibitemOpen
  \bibfield  {author} {\bibinfo {author} {\bibfnamefont {M.}~\bibnamefont
  {Papi\ifmmode~\check{c}\else \v{c}\fi{}}}\ and\ \bibinfo {author}
  {\bibfnamefont {I.}~\bibnamefont {de~Vega}},\ }\bibfield  {title} {\bibinfo
  {title} {Neural-network-based qubit-environment characterization},\ }\href
  {https://doi.org/10.1103/PhysRevA.105.022605} {\bibfield  {journal} {\bibinfo
   {journal} {Phys. Rev. A}\ }\textbf {\bibinfo {volume} {105}},\ \bibinfo
  {pages} {022605} (\bibinfo {year} {2022})}\BibitemShut {NoStop}%
\bibitem [{\citenamefont {Koutn\'y}\ \emph {et~al.}(2022)\citenamefont
  {Koutn\'y}, \citenamefont {Motka}, \citenamefont {Hradil}, \citenamefont
  {\ifmmode \check{R}\else \v{R}\fi{}eh\'a\ifmmode~\check{c}\else
  \v{c}\fi{}ek},\ and\ \citenamefont {S\'anchez-Soto}}]{koutny22}%
  \BibitemOpen
  \bibfield  {author} {\bibinfo {author} {\bibfnamefont {D.}~\bibnamefont
  {Koutn\'y}}, \bibinfo {author} {\bibfnamefont {L.}~\bibnamefont {Motka}},
  \bibinfo {author} {\bibfnamefont {Z.~c.~v.}\ \bibnamefont {Hradil}}, \bibinfo
  {author} {\bibfnamefont {J.}~\bibnamefont {\ifmmode \check{R}\else
  \v{R}\fi{}eh\'a\ifmmode~\check{c}\else \v{c}\fi{}ek}},\ and\ \bibinfo
  {author} {\bibfnamefont {L.~L.}\ \bibnamefont {S\'anchez-Soto}},\ }\bibfield
  {title} {\bibinfo {title} {Neural-network quantum state tomography},\ }\href
  {https://doi.org/10.1103/PhysRevA.106.012409} {\bibfield  {journal} {\bibinfo
   {journal} {Phys. Rev. A}\ }\textbf {\bibinfo {volume} {106}},\ \bibinfo
  {pages} {012409} (\bibinfo {year} {2022})}\BibitemShut {NoStop}%
\bibitem [{\citenamefont {Mavadia}\ \emph {et~al.}(2017)\citenamefont
  {Mavadia}, \citenamefont {Frey}, \citenamefont {Sastrawan}, \citenamefont
  {Dona},\ and\ \citenamefont {Biercuk}}]{mavadia17}%
  \BibitemOpen
  \bibfield  {author} {\bibinfo {author} {\bibfnamefont {S.}~\bibnamefont
  {Mavadia}}, \bibinfo {author} {\bibfnamefont {V.}~\bibnamefont {Frey}},
  \bibinfo {author} {\bibfnamefont {J.}~\bibnamefont {Sastrawan}}, \bibinfo
  {author} {\bibfnamefont {S.}~\bibnamefont {Dona}},\ and\ \bibinfo {author}
  {\bibfnamefont {M.~J.}\ \bibnamefont {Biercuk}},\ }\bibfield  {title}
  {\bibinfo {title} {Prediction and real-time compensation of qubit decoherence
  via machine learning},\ }\href {https://doi.org/10.1038/ncomms14106}
  {\bibfield  {journal} {\bibinfo  {journal} {Nat. Commun.}\ }\textbf {\bibinfo
  {volume} {8}},\ \bibinfo {pages} {14106} (\bibinfo {year}
  {2017})}\BibitemShut {NoStop}%
\bibitem [{\citenamefont {Niu}\ \emph {et~al.}(2019)\citenamefont {Niu},
  \citenamefont {Boixo}, \citenamefont {Smelyanskiy},\ and\ \citenamefont
  {Neven}}]{niu19}%
  \BibitemOpen
  \bibfield  {author} {\bibinfo {author} {\bibfnamefont {M.~Y.}\ \bibnamefont
  {Niu}}, \bibinfo {author} {\bibfnamefont {S.}~\bibnamefont {Boixo}}, \bibinfo
  {author} {\bibfnamefont {V.~N.}\ \bibnamefont {Smelyanskiy}},\ and\ \bibinfo
  {author} {\bibfnamefont {H.}~\bibnamefont {Neven}},\ }\bibfield  {title}
  {\bibinfo {title} {Universal quantum control through deep reinforcement
  learning},\ }\href {https://doi.org/10.1038/s41534-019-0141-3} {\bibfield
  {journal} {\bibinfo  {journal} {npj Quantum Inf.}\ }\textbf {\bibinfo
  {volume} {5}},\ \bibinfo {pages} {33} (\bibinfo {year} {2019})}\BibitemShut
  {NoStop}%
\bibitem [{\citenamefont {Fallani}\ \emph {et~al.}(2022)\citenamefont
  {Fallani}, \citenamefont {Rossi}, \citenamefont {Tamascelli},\ and\
  \citenamefont {Genoni}}]{fallani22}%
  \BibitemOpen
  \bibfield  {author} {\bibinfo {author} {\bibfnamefont {A.}~\bibnamefont
  {Fallani}}, \bibinfo {author} {\bibfnamefont {M.~A.~C.}\ \bibnamefont
  {Rossi}}, \bibinfo {author} {\bibfnamefont {D.}~\bibnamefont {Tamascelli}},\
  and\ \bibinfo {author} {\bibfnamefont {M.~G.}\ \bibnamefont {Genoni}},\
  }\bibfield  {title} {\bibinfo {title} {Learning feedback control strategies
  for quantum metrology},\ }\bibfield  {journal} {\bibinfo  {journal} {PRX
  Quantum}\ }\textbf {\bibinfo {volume} {3}},\ \href
  {https://doi.org/10.1103/PRXQuantum.3.020310} {10.1103/PRXQuantum.3.020310}
  (\bibinfo {year} {2022})\BibitemShut {NoStop}%
\bibitem [{\citenamefont {Huang}\ \emph {et~al.}(2022)\citenamefont {Huang},
  \citenamefont {Ban}, \citenamefont {Sherman},\ and\ \citenamefont
  {Chen}}]{huang22}%
  \BibitemOpen
  \bibfield  {author} {\bibinfo {author} {\bibfnamefont {T.}~\bibnamefont
  {Huang}}, \bibinfo {author} {\bibfnamefont {Y.}~\bibnamefont {Ban}}, \bibinfo
  {author} {\bibfnamefont {E.~Y.}\ \bibnamefont {Sherman}},\ and\ \bibinfo
  {author} {\bibfnamefont {X.}~\bibnamefont {Chen}},\ }\bibfield  {title}
  {\bibinfo {title} {Machine-learning-assisted quantum control in a random
  environment},\ }\href {https://doi.org/10.1103/PhysRevApplied.17.024040}
  {\bibfield  {journal} {\bibinfo  {journal} {Phys. Rev. Applied}\ }\textbf
  {\bibinfo {volume} {17}},\ \bibinfo {pages} {024040} (\bibinfo {year}
  {2022})}\BibitemShut {NoStop}%
\bibitem [{\citenamefont {Porotti}\ \emph {et~al.}(2019)\citenamefont
  {Porotti}, \citenamefont {Tamascelli}, \citenamefont {Restelli},\ and\
  \citenamefont {Prati}}]{porotti19}%
  \BibitemOpen
  \bibfield  {author} {\bibinfo {author} {\bibfnamefont {R.}~\bibnamefont
  {Porotti}}, \bibinfo {author} {\bibfnamefont {D.}~\bibnamefont {Tamascelli}},
  \bibinfo {author} {\bibfnamefont {M.}~\bibnamefont {Restelli}},\ and\
  \bibinfo {author} {\bibfnamefont {E.}~\bibnamefont {Prati}},\ }\bibfield
  {title} {\bibinfo {title} {Coherent transport of quantum states by deep
  reinforcement learning},\ }\href {https://doi.org/10.1038/s42005-019-0169-x}
  {\bibfield  {journal} {\bibinfo  {journal} {Commun. Phys.}\ }\textbf
  {\bibinfo {volume} {2}},\ \bibinfo {pages} {61} (\bibinfo {year}
  {2019})}\BibitemShut {NoStop}%
\bibitem [{\citenamefont {Brown}\ \emph {et~al.}(2021)\citenamefont {Brown},
  \citenamefont {Sgroi}, \citenamefont {Giannelli}, \citenamefont {Paraoanu},
  \citenamefont {Paladino}, \citenamefont {Falci}, \citenamefont
  {Paternostro},\ and\ \citenamefont {Ferraro}}]{Brown_2021}%
  \BibitemOpen
  \bibfield  {author} {\bibinfo {author} {\bibfnamefont {J.}~\bibnamefont
  {Brown}}, \bibinfo {author} {\bibfnamefont {P.}~\bibnamefont {Sgroi}},
  \bibinfo {author} {\bibfnamefont {L.}~\bibnamefont {Giannelli}}, \bibinfo
  {author} {\bibfnamefont {G.~S.}\ \bibnamefont {Paraoanu}}, \bibinfo {author}
  {\bibfnamefont {E.}~\bibnamefont {Paladino}}, \bibinfo {author}
  {\bibfnamefont {G.}~\bibnamefont {Falci}}, \bibinfo {author} {\bibfnamefont
  {M.}~\bibnamefont {Paternostro}},\ and\ \bibinfo {author} {\bibfnamefont
  {A.}~\bibnamefont {Ferraro}},\ }\bibfield  {title} {\bibinfo {title}
  {Reinforcement learning-enhanced protocols for coherent population-transfer
  in three-level quantum systems},\ }\href
  {https://doi.org/10.1088/1367-2630/ac2393} {\bibfield  {journal} {\bibinfo
  {journal} {New J. Phys.}\ }\textbf {\bibinfo {volume} {23}},\ \bibinfo
  {pages} {093035} (\bibinfo {year} {2021})}\BibitemShut {NoStop}%
\bibitem [{\citenamefont {Lumino}\ \emph {et~al.}(2018)\citenamefont {Lumino},
  \citenamefont {Polino}, \citenamefont {Rab}, \citenamefont {Milani},
  \citenamefont {Spagnolo}, \citenamefont {Wiebe},\ and\ \citenamefont
  {Sciarrino}}]{lumino18}%
  \BibitemOpen
  \bibfield  {author} {\bibinfo {author} {\bibfnamefont {A.}~\bibnamefont
  {Lumino}}, \bibinfo {author} {\bibfnamefont {E.}~\bibnamefont {Polino}},
  \bibinfo {author} {\bibfnamefont {A.~S.}\ \bibnamefont {Rab}}, \bibinfo
  {author} {\bibfnamefont {G.}~\bibnamefont {Milani}}, \bibinfo {author}
  {\bibfnamefont {N.}~\bibnamefont {Spagnolo}}, \bibinfo {author}
  {\bibfnamefont {N.}~\bibnamefont {Wiebe}},\ and\ \bibinfo {author}
  {\bibfnamefont {F.}~\bibnamefont {Sciarrino}},\ }\bibfield  {title} {\bibinfo
  {title} {Experimental phase estimation enhanced by machine learning},\ }\href
  {https://doi.org/10.1103/PhysRevApplied.10.044033} {\bibfield  {journal}
  {\bibinfo  {journal} {Phys. Rev. Applied}\ }\textbf {\bibinfo {volume}
  {10}},\ \bibinfo {pages} {044033} (\bibinfo {year} {2018})}\BibitemShut
  {NoStop}%
\bibitem [{\citenamefont {Cimini}\ \emph {et~al.}(2019)\citenamefont {Cimini},
  \citenamefont {Gianani}, \citenamefont {Spagnolo}, \citenamefont {Leccese},
  \citenamefont {Sciarrino},\ and\ \citenamefont {Barbieri}}]{cimini2019}%
  \BibitemOpen
  \bibfield  {author} {\bibinfo {author} {\bibfnamefont {V.}~\bibnamefont
  {Cimini}}, \bibinfo {author} {\bibfnamefont {I.}~\bibnamefont {Gianani}},
  \bibinfo {author} {\bibfnamefont {N.}~\bibnamefont {Spagnolo}}, \bibinfo
  {author} {\bibfnamefont {F.}~\bibnamefont {Leccese}}, \bibinfo {author}
  {\bibfnamefont {F.}~\bibnamefont {Sciarrino}},\ and\ \bibinfo {author}
  {\bibfnamefont {M.}~\bibnamefont {Barbieri}},\ }\bibfield  {title} {\bibinfo
  {title} {Calibration of quantum sensors by neural networks},\ }\href
  {https://doi.org/10.1103/PhysRevLett.123.230502} {\bibfield  {journal}
  {\bibinfo  {journal} {Phys. Rev. Lett.}\ }\textbf {\bibinfo {volume} {123}},\
  \bibinfo {pages} {230502} (\bibinfo {year} {2019})}\BibitemShut {NoStop}%
\bibitem [{\citenamefont {Ming}\ \emph {et~al.}(2019)\citenamefont {Ming},
  \citenamefont {Lin}, \citenamefont {Bartlett},\ and\ \citenamefont
  {Zhang}}]{ming19}%
  \BibitemOpen
  \bibfield  {author} {\bibinfo {author} {\bibfnamefont {Y.}~\bibnamefont
  {Ming}}, \bibinfo {author} {\bibfnamefont {C.-T.}\ \bibnamefont {Lin}},
  \bibinfo {author} {\bibfnamefont {S.~D.}\ \bibnamefont {Bartlett}},\ and\
  \bibinfo {author} {\bibfnamefont {W.-W.}\ \bibnamefont {Zhang}},\ }\bibfield
  {title} {\bibinfo {title} {Quantum topology identification with deep neural
  networks and quantum walks},\ }\href
  {https://doi.org/10.1038/s41524-019-0224-x} {\bibfield  {journal} {\bibinfo
  {journal} {npj Comput. Mater.}\ }\textbf {\bibinfo {volume} {5}},\ \bibinfo
  {pages} {88} (\bibinfo {year} {2019})}\BibitemShut {NoStop}%
\bibitem [{\citenamefont {Valeri}\ \emph {et~al.}(2020)\citenamefont {Valeri},
  \citenamefont {Polino}, \citenamefont {Poderini}, \citenamefont {Gianani},
  \citenamefont {Corrielli}, \citenamefont {Crespi}, \citenamefont {Osellame},
  \citenamefont {Spagnolo},\ and\ \citenamefont {Sciarrino}}]{Valeri2020}%
  \BibitemOpen
  \bibfield  {author} {\bibinfo {author} {\bibfnamefont {M.}~\bibnamefont
  {Valeri}}, \bibinfo {author} {\bibfnamefont {E.}~\bibnamefont {Polino}},
  \bibinfo {author} {\bibfnamefont {D.}~\bibnamefont {Poderini}}, \bibinfo
  {author} {\bibfnamefont {I.}~\bibnamefont {Gianani}}, \bibinfo {author}
  {\bibfnamefont {G.}~\bibnamefont {Corrielli}}, \bibinfo {author}
  {\bibfnamefont {A.}~\bibnamefont {Crespi}}, \bibinfo {author} {\bibfnamefont
  {R.}~\bibnamefont {Osellame}}, \bibinfo {author} {\bibfnamefont
  {N.}~\bibnamefont {Spagnolo}},\ and\ \bibinfo {author} {\bibfnamefont
  {F.}~\bibnamefont {Sciarrino}},\ }\bibfield  {title} {\bibinfo {title}
  {{Experimental adaptive Bayesian estimation of multiple phases with limited
  data}},\ }\bibfield  {journal} {\bibinfo  {journal} {npj Quantum Inf.}\
  }\textbf {\bibinfo {volume} {6}},\ \href
  {https://doi.org/10.1038/s41534-020-00326-6} {10.1038/s41534-020-00326-6}
  (\bibinfo {year} {2020})\BibitemShut {NoStop}%
\bibitem [{\citenamefont {Cimini}\ \emph {et~al.}(2021)\citenamefont {Cimini},
  \citenamefont {Polino}, \citenamefont {Valeri}, \citenamefont {Gianani},
  \citenamefont {Spagnolo}, \citenamefont {Corrielli}, \citenamefont {Crespi},
  \citenamefont {Osellame}, \citenamefont {Barbieri},\ and\ \citenamefont
  {Sciarrino}}]{cimini2021}%
  \BibitemOpen
  \bibfield  {author} {\bibinfo {author} {\bibfnamefont {V.}~\bibnamefont
  {Cimini}}, \bibinfo {author} {\bibfnamefont {E.}~\bibnamefont {Polino}},
  \bibinfo {author} {\bibfnamefont {M.}~\bibnamefont {Valeri}}, \bibinfo
  {author} {\bibfnamefont {I.}~\bibnamefont {Gianani}}, \bibinfo {author}
  {\bibfnamefont {N.}~\bibnamefont {Spagnolo}}, \bibinfo {author}
  {\bibfnamefont {G.}~\bibnamefont {Corrielli}}, \bibinfo {author}
  {\bibfnamefont {A.}~\bibnamefont {Crespi}}, \bibinfo {author} {\bibfnamefont
  {R.}~\bibnamefont {Osellame}}, \bibinfo {author} {\bibfnamefont
  {M.}~\bibnamefont {Barbieri}},\ and\ \bibinfo {author} {\bibfnamefont
  {F.}~\bibnamefont {Sciarrino}},\ }\bibfield  {title} {\bibinfo {title}
  {Calibration of multiparameter sensors via machine learning at the
  single-photon level},\ }\href
  {https://doi.org/10.1103/PhysRevApplied.15.044003} {\bibfield  {journal}
  {\bibinfo  {journal} {Phys. Rev. Applied}\ }\textbf {\bibinfo {volume}
  {15}},\ \bibinfo {pages} {044003} (\bibinfo {year} {2021})}\BibitemShut
  {NoStop}%
\bibitem [{\citenamefont {Ban}\ \emph {et~al.}(2021)\citenamefont {Ban},
  \citenamefont {Echanobe}, \citenamefont {Ding}, \citenamefont {Puebla},\ and\
  \citenamefont {Casanova}}]{Ban21}%
  \BibitemOpen
  \bibfield  {author} {\bibinfo {author} {\bibfnamefont {Y.}~\bibnamefont
  {Ban}}, \bibinfo {author} {\bibfnamefont {J.}~\bibnamefont {Echanobe}},
  \bibinfo {author} {\bibfnamefont {Y.}~\bibnamefont {Ding}}, \bibinfo {author}
  {\bibfnamefont {R.}~\bibnamefont {Puebla}},\ and\ \bibinfo {author}
  {\bibfnamefont {J.}~\bibnamefont {Casanova}},\ }\bibfield  {title} {\bibinfo
  {title} {Neural-network-based parameter estimation for quantum detection},\
  }\href {https://doi.org/10.1088/2058-9565/ac16ed} {\bibfield  {journal}
  {\bibinfo  {journal} {Quantum Sci. Technol.}\ }\textbf {\bibinfo {volume}
  {6}},\ \bibinfo {pages} {045012} (\bibinfo {year} {2021})}\BibitemShut
  {NoStop}%
\bibitem [{\citenamefont {Palmieri}\ \emph {et~al.}(2021)\citenamefont
  {Palmieri}, \citenamefont {Bianchi}, \citenamefont {Paris},\ and\
  \citenamefont {Benedetti}}]{palmieri21}%
  \BibitemOpen
  \bibfield  {author} {\bibinfo {author} {\bibfnamefont {A.~M.}\ \bibnamefont
  {Palmieri}}, \bibinfo {author} {\bibfnamefont {F.}~\bibnamefont {Bianchi}},
  \bibinfo {author} {\bibfnamefont {M.~G.~A.}\ \bibnamefont {Paris}},\ and\
  \bibinfo {author} {\bibfnamefont {C.}~\bibnamefont {Benedetti}},\ }\bibfield
  {title} {\bibinfo {title} {Multiclass classification of dephasing channels},\
  }\href {https://doi.org/10.1103/PhysRevA.104.052412} {\bibfield  {journal}
  {\bibinfo  {journal} {Phys. Rev. A}\ }\textbf {\bibinfo {volume} {104}},\
  \bibinfo {pages} {052412} (\bibinfo {year} {2021})}\BibitemShut {NoStop}%
\bibitem [{\citenamefont {Gianani}\ \emph {et~al.}(2022)\citenamefont
  {Gianani}, \citenamefont {Mastroserio}, \citenamefont {Buffoni},
  \citenamefont {Bruno}, \citenamefont {Donati}, \citenamefont {Cimini},
  \citenamefont {Barbieri}, \citenamefont {Cataliotti},\ and\ \citenamefont
  {Caruso}}]{gianani22}%
  \BibitemOpen
  \bibfield  {author} {\bibinfo {author} {\bibfnamefont {I.}~\bibnamefont
  {Gianani}}, \bibinfo {author} {\bibfnamefont {I.}~\bibnamefont
  {Mastroserio}}, \bibinfo {author} {\bibfnamefont {L.}~\bibnamefont
  {Buffoni}}, \bibinfo {author} {\bibfnamefont {N.}~\bibnamefont {Bruno}},
  \bibinfo {author} {\bibfnamefont {L.}~\bibnamefont {Donati}}, \bibinfo
  {author} {\bibfnamefont {V.}~\bibnamefont {Cimini}}, \bibinfo {author}
  {\bibfnamefont {M.}~\bibnamefont {Barbieri}}, \bibinfo {author}
  {\bibfnamefont {F.~S.}\ \bibnamefont {Cataliotti}},\ and\ \bibinfo {author}
  {\bibfnamefont {F.}~\bibnamefont {Caruso}},\ }\bibfield  {title} {\bibinfo
  {title} {Experimental quantum embedding for machine learning},\ }\href
  {https://doi.org/https://doi.org/10.1002/qute.202100140} {\bibfield
  {journal} {\bibinfo  {journal} {Adv. Quantum Technol.}\ }\textbf {\bibinfo
  {volume} {5}},\ \bibinfo {pages} {2100140} (\bibinfo {year}
  {2022})}\BibitemShut {NoStop}%
\bibitem [{\citenamefont {Granade}\ \emph {et~al.}(2012)\citenamefont
  {Granade}, \citenamefont {Ferrie}, \citenamefont {Wiebe},\ and\ \citenamefont
  {Cory}}]{granade12}%
  \BibitemOpen
  \bibfield  {author} {\bibinfo {author} {\bibfnamefont {C.~E.}\ \bibnamefont
  {Granade}}, \bibinfo {author} {\bibfnamefont {C.}~\bibnamefont {Ferrie}},
  \bibinfo {author} {\bibfnamefont {N.}~\bibnamefont {Wiebe}},\ and\ \bibinfo
  {author} {\bibfnamefont {D.~G.}\ \bibnamefont {Cory}},\ }\bibfield  {title}
  {\bibinfo {title} {Robust online hamiltonian learning},\ }\href
  {https://doi.org/10.1088/1367-2630/14/10/103013} {\bibfield  {journal}
  {\bibinfo  {journal} {New J. Phys.}\ }\textbf {\bibinfo {volume} {14}},\
  \bibinfo {pages} {103013} (\bibinfo {year} {2012})}\BibitemShut {NoStop}%
\bibitem [{\citenamefont {Wiebe}\ \emph
  {et~al.}(2014{\natexlab{a}})\citenamefont {Wiebe}, \citenamefont {Granade},
  \citenamefont {Ferrie},\ and\ \citenamefont {Cory}}]{wiebe14}%
  \BibitemOpen
  \bibfield  {author} {\bibinfo {author} {\bibfnamefont {N.}~\bibnamefont
  {Wiebe}}, \bibinfo {author} {\bibfnamefont {C.}~\bibnamefont {Granade}},
  \bibinfo {author} {\bibfnamefont {C.}~\bibnamefont {Ferrie}},\ and\ \bibinfo
  {author} {\bibfnamefont {D.~G.}\ \bibnamefont {Cory}},\ }\bibfield  {title}
  {\bibinfo {title} {Hamiltonian learning and certification using quantum
  resources},\ }\href {https://doi.org/10.1103/PhysRevLett.112.190501}
  {\bibfield  {journal} {\bibinfo  {journal} {Phys. Rev. Lett.}\ }\textbf
  {\bibinfo {volume} {112}},\ \bibinfo {pages} {190501} (\bibinfo {year}
  {2014}{\natexlab{a}})}\BibitemShut {NoStop}%
\bibitem [{\citenamefont {Wiebe}\ \emph
  {et~al.}(2014{\natexlab{b}})\citenamefont {Wiebe}, \citenamefont {Granade},
  \citenamefont {Ferrie},\ and\ \citenamefont {Cory}}]{wiebe14A}%
  \BibitemOpen
  \bibfield  {author} {\bibinfo {author} {\bibfnamefont {N.}~\bibnamefont
  {Wiebe}}, \bibinfo {author} {\bibfnamefont {C.}~\bibnamefont {Granade}},
  \bibinfo {author} {\bibfnamefont {C.}~\bibnamefont {Ferrie}},\ and\ \bibinfo
  {author} {\bibfnamefont {D.}~\bibnamefont {Cory}},\ }\bibfield  {title}
  {\bibinfo {title} {Quantum hamiltonian learning using imperfect quantum
  resources},\ }\href {https://doi.org/10.1103/PhysRevA.89.042314} {\bibfield
  {journal} {\bibinfo  {journal} {Phys. Rev. A}\ }\textbf {\bibinfo {volume}
  {89}},\ \bibinfo {pages} {042314} (\bibinfo {year}
  {2014}{\natexlab{b}})}\BibitemShut {NoStop}%
\bibitem [{\citenamefont {Cao}\ \emph {et~al.}(2020)\citenamefont {Cao},
  \citenamefont {Hou}, \citenamefont {Cao},\ and\ \citenamefont
  {Zeng}}]{Cao20}%
  \BibitemOpen
  \bibfield  {author} {\bibinfo {author} {\bibfnamefont {C.}~\bibnamefont
  {Cao}}, \bibinfo {author} {\bibfnamefont {S.-Y.}\ \bibnamefont {Hou}},
  \bibinfo {author} {\bibfnamefont {N.}~\bibnamefont {Cao}},\ and\ \bibinfo
  {author} {\bibfnamefont {B.}~\bibnamefont {Zeng}},\ }\bibfield  {title}
  {\bibinfo {title} {Supervised learning in hamiltonian reconstruction from
  local measurements on eigenstates},\ }\href
  {https://doi.org/10.1088/1361-648X/abc4cf} {\bibfield  {journal} {\bibinfo
  {journal} {J. Phys.: Cond. Matt.}\ }\textbf {\bibinfo {volume} {33}},\
  \bibinfo {pages} {064002} (\bibinfo {year} {2020})}\BibitemShut {NoStop}%
\bibitem [{\citenamefont {Che}\ \emph {et~al.}(2021)\citenamefont {Che},
  \citenamefont {Wei}, \citenamefont {Huang}, \citenamefont {Zhao},
  \citenamefont {Xue}, \citenamefont {Nie}, \citenamefont {Li}, \citenamefont
  {Lu},\ and\ \citenamefont {Xin}}]{che21}%
  \BibitemOpen
  \bibfield  {author} {\bibinfo {author} {\bibfnamefont {L.}~\bibnamefont
  {Che}}, \bibinfo {author} {\bibfnamefont {C.}~\bibnamefont {Wei}}, \bibinfo
  {author} {\bibfnamefont {Y.}~\bibnamefont {Huang}}, \bibinfo {author}
  {\bibfnamefont {D.}~\bibnamefont {Zhao}}, \bibinfo {author} {\bibfnamefont
  {S.}~\bibnamefont {Xue}}, \bibinfo {author} {\bibfnamefont {X.}~\bibnamefont
  {Nie}}, \bibinfo {author} {\bibfnamefont {J.}~\bibnamefont {Li}}, \bibinfo
  {author} {\bibfnamefont {D.}~\bibnamefont {Lu}},\ and\ \bibinfo {author}
  {\bibfnamefont {T.}~\bibnamefont {Xin}},\ }\bibfield  {title} {\bibinfo
  {title} {Learning quantum hamiltonians from single-qubit measurements},\
  }\href {https://doi.org/10.1103/PhysRevResearch.3.023246} {\bibfield
  {journal} {\bibinfo  {journal} {Phys. Rev. Research}\ }\textbf {\bibinfo
  {volume} {3}},\ \bibinfo {pages} {023246} (\bibinfo {year}
  {2021})}\BibitemShut {NoStop}%
\bibitem [{\citenamefont {Rattacaso}\ \emph {et~al.}(2022)\citenamefont
  {Rattacaso}, \citenamefont {Passarelli},\ and\ \citenamefont
  {Lucignano}}]{rattacaso22}%
  \BibitemOpen
  \bibfield  {author} {\bibinfo {author} {\bibfnamefont {D.}~\bibnamefont
  {Rattacaso}}, \bibinfo {author} {\bibfnamefont {G.}~\bibnamefont
  {Passarelli}},\ and\ \bibinfo {author} {\bibfnamefont {P.}~\bibnamefont
  {Lucignano}},\ }\bibfield  {title} {\bibinfo {title} {High-accuracy
  hamiltonian learning via delocalized quantum state evolutions},\ }\bibfield
  {journal} {\bibinfo  {journal} {{arXiv.2204.03997}}\ }\href
  {https://doi.org/10.48550/arXiv.2204.03997} {10.48550/arXiv.2204.03997}
  (\bibinfo {year} {2022})\BibitemShut {NoStop}%
\bibitem [{\citenamefont {Wang}\ \emph {et~al.}(2017)\citenamefont {Wang},
  \citenamefont {Paesani}, \citenamefont {Santagati}, \citenamefont {Knauer},
  \citenamefont {Gentile}, \citenamefont {Wiebe}, \citenamefont {Petruzzella},
  \citenamefont {O'Brien}, \citenamefont {Rarity}, \citenamefont {Laing},\ and\
  \citenamefont {Thompson}}]{jianwei17}%
  \BibitemOpen
  \bibfield  {author} {\bibinfo {author} {\bibfnamefont {J.}~\bibnamefont
  {Wang}}, \bibinfo {author} {\bibfnamefont {S.}~\bibnamefont {Paesani}},
  \bibinfo {author} {\bibfnamefont {R.}~\bibnamefont {Santagati}}, \bibinfo
  {author} {\bibfnamefont {S.}~\bibnamefont {Knauer}}, \bibinfo {author}
  {\bibfnamefont {A.~A.}\ \bibnamefont {Gentile}}, \bibinfo {author}
  {\bibfnamefont {N.}~\bibnamefont {Wiebe}}, \bibinfo {author} {\bibfnamefont
  {M.}~\bibnamefont {Petruzzella}}, \bibinfo {author} {\bibfnamefont {J.~L.}\
  \bibnamefont {O'Brien}}, \bibinfo {author} {\bibfnamefont {J.~G.}\
  \bibnamefont {Rarity}}, \bibinfo {author} {\bibfnamefont {A.}~\bibnamefont
  {Laing}},\ and\ \bibinfo {author} {\bibfnamefont {M.~G.}\ \bibnamefont
  {Thompson}},\ }\bibfield  {title} {\bibinfo {title} {Experimental quantum
  hamiltonian learning},\ }\href {https://doi.org/10.1038/nphys4074} {\bibfield
   {journal} {\bibinfo  {journal} {Nat. Phys.}\ }\textbf {\bibinfo {volume}
  {13}},\ \bibinfo {pages} {551} (\bibinfo {year} {2017})}\BibitemShut
  {NoStop}%
\bibitem [{\citenamefont {Gentile}\ \emph {et~al.}(2021)\citenamefont
  {Gentile}, \citenamefont {Flynn}, \citenamefont {Knauer}, \citenamefont
  {Wiebe}, \citenamefont {Paesani}, \citenamefont {Granade}, \citenamefont
  {Rarity}, \citenamefont {Santagati},\ and\ \citenamefont
  {Laing}}]{gentile21}%
  \BibitemOpen
  \bibfield  {author} {\bibinfo {author} {\bibfnamefont {A.~A.}\ \bibnamefont
  {Gentile}}, \bibinfo {author} {\bibfnamefont {B.}~\bibnamefont {Flynn}},
  \bibinfo {author} {\bibfnamefont {S.}~\bibnamefont {Knauer}}, \bibinfo
  {author} {\bibfnamefont {N.}~\bibnamefont {Wiebe}}, \bibinfo {author}
  {\bibfnamefont {S.}~\bibnamefont {Paesani}}, \bibinfo {author} {\bibfnamefont
  {C.~E.}\ \bibnamefont {Granade}}, \bibinfo {author} {\bibfnamefont {J.~G.}\
  \bibnamefont {Rarity}}, \bibinfo {author} {\bibfnamefont {R.}~\bibnamefont
  {Santagati}},\ and\ \bibinfo {author} {\bibfnamefont {A.}~\bibnamefont
  {Laing}},\ }\bibfield  {title} {\bibinfo {title} {Learning models of quantum
  systems from experiments},\ }\href
  {https://doi.org/10.1038/s41567-021-01201-7} {\bibfield  {journal} {\bibinfo
  {journal} {Nat. Phys.}\ }\textbf {\bibinfo {volume} {17}},\ \bibinfo {pages}
  {837} (\bibinfo {year} {2021})}\BibitemShut {NoStop}%
\bibitem [{\citenamefont {Giovannetti}\ \emph {et~al.}(2004)\citenamefont
  {Giovannetti}, \citenamefont {Lloyd},\ and\ \citenamefont
  {Maccone}}]{Giovannetti2004}%
  \BibitemOpen
  \bibfield  {author} {\bibinfo {author} {\bibfnamefont {V.}~\bibnamefont
  {Giovannetti}}, \bibinfo {author} {\bibfnamefont {S.}~\bibnamefont {Lloyd}},\
  and\ \bibinfo {author} {\bibfnamefont {L.}~\bibnamefont {Maccone}},\
  }\bibfield  {title} {\bibinfo {title} {{Quantum-Enhanced Measurements :
  Beating the Standard Quantum Limit}},\ }\href@noop {} {\bibfield  {journal}
  {\bibinfo  {journal} {Science}\ }\textbf {\bibinfo {volume} {306}},\ \bibinfo
  {pages} {1330} (\bibinfo {year} {2004})}\BibitemShut {NoStop}%
\bibitem [{\citenamefont {Giovannetti}\ \emph {et~al.}(2011)\citenamefont
  {Giovannetti}, \citenamefont {Lloyd},\ and\ \citenamefont
  {Maccone}}]{Giovannetti2011}%
  \BibitemOpen
  \bibfield  {author} {\bibinfo {author} {\bibfnamefont {V.}~\bibnamefont
  {Giovannetti}}, \bibinfo {author} {\bibfnamefont {S.}~\bibnamefont {Lloyd}},\
  and\ \bibinfo {author} {\bibfnamefont {L.}~\bibnamefont {Maccone}},\
  }\bibfield  {title} {\bibinfo {title} {{Advances in quantum metrology}},\
  }\href {https://doi.org/10.1038/nphoton.2011.35} {\bibfield  {journal}
  {\bibinfo  {journal} {Nat. Photon.}\ }\textbf {\bibinfo {volume} {5}},\
  \bibinfo {pages} {222} (\bibinfo {year} {2011})}\BibitemShut {NoStop}%
\bibitem [{\citenamefont {Paris}(2009)}]{PARIS2009}%
  \BibitemOpen
  \bibfield  {author} {\bibinfo {author} {\bibfnamefont {M.~G.~A.}\
  \bibnamefont {Paris}},\ }\bibfield  {title} {\bibinfo {title} {{Quantum
  estimation for quantum technology}},\ }\href
  {https://doi.org/10.1142/S0219749909004839} {\bibfield  {journal} {\bibinfo
  {journal} {Int. J. Quantum Inf.}\ }\textbf {\bibinfo {volume} {07}},\
  \bibinfo {pages} {125} (\bibinfo {year} {2009})}\BibitemShut {NoStop}%
\bibitem [{\citenamefont {Albarelli}\ \emph {et~al.}(2020)\citenamefont
  {Albarelli}, \citenamefont {Barbieri}, \citenamefont {Genoni},\ and\
  \citenamefont {Gianani}}]{albarelli20}%
  \BibitemOpen
  \bibfield  {author} {\bibinfo {author} {\bibfnamefont {F.}~\bibnamefont
  {Albarelli}}, \bibinfo {author} {\bibfnamefont {M.}~\bibnamefont {Barbieri}},
  \bibinfo {author} {\bibfnamefont {M.}~\bibnamefont {Genoni}},\ and\ \bibinfo
  {author} {\bibfnamefont {I.}~\bibnamefont {Gianani}},\ }\bibfield  {title}
  {\bibinfo {title} {A perspective on multiparameter quantum metrology: From
  theoretical tools to applications in quantum imaging},\ }\href
  {https://doi.org/https://doi.org/10.1016/j.physleta.2020.126311} {\bibfield
  {journal} {\bibinfo  {journal} {Phys. Lett. A}\ }\textbf {\bibinfo {volume}
  {384}},\ \bibinfo {pages} {126311} (\bibinfo {year} {2020})}\BibitemShut
  {NoStop}%
\bibitem [{\citenamefont {Plenio}\ and\ \citenamefont
  {Huelga}(2008)}]{Plenio_2008}%
  \BibitemOpen
  \bibfield  {author} {\bibinfo {author} {\bibfnamefont {M.~B.}\ \bibnamefont
  {Plenio}}\ and\ \bibinfo {author} {\bibfnamefont {S.~F.}\ \bibnamefont
  {Huelga}},\ }\bibfield  {title} {\bibinfo {title} {Dephasing-assisted
  transport: quantum networks and biomolecules},\ }\href
  {https://doi.org/10.1088/1367-2630/10/11/113019} {\bibfield  {journal}
  {\bibinfo  {journal} {New J. Phys.}\ }\textbf {\bibinfo {volume} {10}},\
  \bibinfo {pages} {113019} (\bibinfo {year} {2008})}\BibitemShut {NoStop}%
\bibitem [{\citenamefont {M{\"u}lken}\ and\ \citenamefont
  {Blumen}(2011)}]{mulken2011}%
  \BibitemOpen
  \bibfield  {author} {\bibinfo {author} {\bibfnamefont {O.}~\bibnamefont
  {M{\"u}lken}}\ and\ \bibinfo {author} {\bibfnamefont {A.}~\bibnamefont
  {Blumen}},\ }\bibfield  {title} {\bibinfo {title} {Continuous-time quantum
  walks: Models for coherent transport on complex networks},\ }\href
  {https://doi.org/10.1016/j.physrep.2011.01.002} {\bibfield  {journal}
  {\bibinfo  {journal} {Phys. Rep.}\ }\textbf {\bibinfo {volume} {502}},\
  \bibinfo {pages} {37} (\bibinfo {year} {2011})}\BibitemShut {NoStop}%
\bibitem [{\citenamefont {Uchiyama}\ \emph {et~al.}(2018)\citenamefont
  {Uchiyama}, \citenamefont {Munro},\ and\ \citenamefont
  {Nemoto}}]{uchiyama18}%
  \BibitemOpen
  \bibfield  {author} {\bibinfo {author} {\bibfnamefont {C.}~\bibnamefont
  {Uchiyama}}, \bibinfo {author} {\bibfnamefont {W.~J.}\ \bibnamefont
  {Munro}},\ and\ \bibinfo {author} {\bibfnamefont {K.}~\bibnamefont
  {Nemoto}},\ }\bibfield  {title} {\bibinfo {title} {Environmental engineering
  for quantum energy transport},\ }\href
  {https://doi.org/10.1038/s41534-018-0079-x} {\bibfield  {journal} {\bibinfo
  {journal} {npj Quantum Inf.}\ }\textbf {\bibinfo {volume} {4}},\ \bibinfo
  {pages} {33} (\bibinfo {year} {2018})}\BibitemShut {NoStop}%
\bibitem [{\citenamefont {Childs}\ and\ \citenamefont
  {Goldstone}(2004)}]{childs04}%
  \BibitemOpen
  \bibfield  {author} {\bibinfo {author} {\bibfnamefont {A.~M.}\ \bibnamefont
  {Childs}}\ and\ \bibinfo {author} {\bibfnamefont {J.}~\bibnamefont
  {Goldstone}},\ }\bibfield  {title} {\bibinfo {title} {Spatial search by
  quantum walk},\ }\href {https://doi.org/10.1103/PhysRevA.70.022314}
  {\bibfield  {journal} {\bibinfo  {journal} {Phys. Rev. A}\ }\textbf {\bibinfo
  {volume} {70}},\ \bibinfo {pages} {022314} (\bibinfo {year}
  {2004})}\BibitemShut {NoStop}%
\bibitem [{\citenamefont {Gamble}\ \emph {et~al.}(2010)\citenamefont {Gamble},
  \citenamefont {Friesen}, \citenamefont {Zhou}, \citenamefont {Joynt},\ and\
  \citenamefont {Coppersmith}}]{gamble10}%
  \BibitemOpen
  \bibfield  {author} {\bibinfo {author} {\bibfnamefont {J.~K.}\ \bibnamefont
  {Gamble}}, \bibinfo {author} {\bibfnamefont {M.}~\bibnamefont {Friesen}},
  \bibinfo {author} {\bibfnamefont {D.}~\bibnamefont {Zhou}}, \bibinfo {author}
  {\bibfnamefont {R.}~\bibnamefont {Joynt}},\ and\ \bibinfo {author}
  {\bibfnamefont {S.~N.}\ \bibnamefont {Coppersmith}},\ }\bibfield  {title}
  {\bibinfo {title} {Two-particle quantum walks applied to the graph
  isomorphism problem},\ }\href {https://doi.org/10.1103/PhysRevA.81.052313}
  {\bibfield  {journal} {\bibinfo  {journal} {Phys. Rev. A}\ }\textbf {\bibinfo
  {volume} {81}},\ \bibinfo {pages} {052313} (\bibinfo {year}
  {2010})}\BibitemShut {NoStop}%
\bibitem [{\citenamefont {Chakraborty}\ \emph {et~al.}(2020)\citenamefont
  {Chakraborty}, \citenamefont {Novo},\ and\ \citenamefont
  {Roland}}]{Chakraborty20}%
  \BibitemOpen
  \bibfield  {author} {\bibinfo {author} {\bibfnamefont {S.}~\bibnamefont
  {Chakraborty}}, \bibinfo {author} {\bibfnamefont {L.}~\bibnamefont {Novo}},\
  and\ \bibinfo {author} {\bibfnamefont {J.}~\bibnamefont {Roland}},\
  }\bibfield  {title} {\bibinfo {title} {Optimality of spatial search via
  continuous-time quantum walks},\ }\href
  {https://doi.org/10.1103/PhysRevA.102.032214} {\bibfield  {journal} {\bibinfo
   {journal} {Phys. Rev. A}\ }\textbf {\bibinfo {volume} {102}},\ \bibinfo
  {pages} {032214} (\bibinfo {year} {2020})}\BibitemShut {NoStop}%
\bibitem [{\citenamefont {Atia}\ and\ \citenamefont
  {Chakraborty}(2021)}]{atia21}%
  \BibitemOpen
  \bibfield  {author} {\bibinfo {author} {\bibfnamefont {Y.}~\bibnamefont
  {Atia}}\ and\ \bibinfo {author} {\bibfnamefont {S.}~\bibnamefont
  {Chakraborty}},\ }\bibfield  {title} {\bibinfo {title} {Improved upper bounds
  for the hitting times of quantum walks},\ }\href
  {https://doi.org/10.1103/PhysRevA.104.032215} {\bibfield  {journal} {\bibinfo
   {journal} {Phys. Rev. A}\ }\textbf {\bibinfo {volume} {104}},\ \bibinfo
  {pages} {032215} (\bibinfo {year} {2021})}\BibitemShut {NoStop}%
\bibitem [{\citenamefont {Paris}\ \emph {et~al.}(2021)\citenamefont {Paris},
  \citenamefont {Benedetti},\ and\ \citenamefont {Olivares}}]{paris21}%
  \BibitemOpen
  \bibfield  {author} {\bibinfo {author} {\bibfnamefont {M.~G.~A.}\
  \bibnamefont {Paris}}, \bibinfo {author} {\bibfnamefont {C.}~\bibnamefont
  {Benedetti}},\ and\ \bibinfo {author} {\bibfnamefont {S.}~\bibnamefont
  {Olivares}},\ }\bibfield  {title} {\bibinfo {title} {Improving quantum search
  on simple graphs by pretty good structured oracles},\ }\bibfield  {journal}
  {\bibinfo  {journal} {Symmetry}\ }\textbf {\bibinfo {volume} {13}},\ \href
  {https://doi.org/10.3390/sym13010096} {10.3390/sym13010096} (\bibinfo {year}
  {2021})\BibitemShut {NoStop}%
\bibitem [{\citenamefont {Candeloro}\ \emph {et~al.}(2022)\citenamefont
  {Candeloro}, \citenamefont {Benedetti}, \citenamefont {Genoni},\ and\
  \citenamefont {Paris}}]{candeloro22}%
  \BibitemOpen
  \bibfield  {author} {\bibinfo {author} {\bibfnamefont {A.}~\bibnamefont
  {Candeloro}}, \bibinfo {author} {\bibfnamefont {C.}~\bibnamefont
  {Benedetti}}, \bibinfo {author} {\bibfnamefont {M.~G.}\ \bibnamefont
  {Genoni}},\ and\ \bibinfo {author} {\bibfnamefont {M.~G.~A.}\ \bibnamefont
  {Paris}},\ }\bibfield  {title} {\bibinfo {title} {Feedback-assisted quantum
  search by continuous-time quantum walks},\ }\href
  {https://doi.org/https://doi.org/10.1002/qute.202200093} {\bibfield
  {journal} {\bibinfo  {journal} {Adv. Quantum Technol.}\ ,\ \bibinfo {pages}
  {2200093}} (\bibinfo {year} {2022})}\BibitemShut {NoStop}%
\bibitem [{\citenamefont {Apers}\ \emph {et~al.}(2022)\citenamefont {Apers},
  \citenamefont {Chakraborty}, \citenamefont {Novo},\ and\ \citenamefont
  {Roland}}]{apers22}%
  \BibitemOpen
  \bibfield  {author} {\bibinfo {author} {\bibfnamefont {S.}~\bibnamefont
  {Apers}}, \bibinfo {author} {\bibfnamefont {S.}~\bibnamefont {Chakraborty}},
  \bibinfo {author} {\bibfnamefont {L.}~\bibnamefont {Novo}},\ and\ \bibinfo
  {author} {\bibfnamefont {J.}~\bibnamefont {Roland}},\ }\bibfield  {title}
  {\bibinfo {title} {Quadratic speedup for spatial search by continuous-time
  quantum walk},\ }\href {https://doi.org/10.1103/PhysRevLett.129.160502}
  {\bibfield  {journal} {\bibinfo  {journal} {Phys. Rev. Lett.}\ }\textbf
  {\bibinfo {volume} {129}},\ \bibinfo {pages} {160502} (\bibinfo {year}
  {2022})}\BibitemShut {NoStop}%
\bibitem [{\citenamefont {Childs}(2009)}]{childs09}%
  \BibitemOpen
  \bibfield  {author} {\bibinfo {author} {\bibfnamefont {A.~M.}\ \bibnamefont
  {Childs}},\ }\bibfield  {title} {\bibinfo {title} {Universal computation by
  quantum walk},\ }\href {https://doi.org/10.1103/PhysRevLett.102.180501}
  {\bibfield  {journal} {\bibinfo  {journal} {Phys. Rev. Lett.}\ }\textbf
  {\bibinfo {volume} {102}},\ \bibinfo {pages} {180501} (\bibinfo {year}
  {2009})}\BibitemShut {NoStop}%
\bibitem [{\citenamefont {Lovett}\ \emph {et~al.}(2010)\citenamefont {Lovett},
  \citenamefont {Cooper}, \citenamefont {Everitt}, \citenamefont {Trevers}, ,\
  and\ \citenamefont {Kendon}}]{lovett10}%
  \BibitemOpen
  \bibfield  {author} {\bibinfo {author} {\bibfnamefont {N.~B.}\ \bibnamefont
  {Lovett}}, \bibinfo {author} {\bibfnamefont {S.}~\bibnamefont {Cooper}},
  \bibinfo {author} {\bibfnamefont {M.}~\bibnamefont {Everitt}}, \bibinfo
  {author} {\bibfnamefont {M.}~\bibnamefont {Trevers}}, ,\ and\ \bibinfo
  {author} {\bibfnamefont {V.}~\bibnamefont {Kendon}},\ }\bibfield  {title}
  {\bibinfo {title} {Universal quantum computation using the discrete-time
  quantum walk},\ }\href
  {https://doi.org/https://doi.org/10.1103/PhysRevA.81.042330} {\bibfield
  {journal} {\bibinfo  {journal} {Phys. Rev. A}\ }\textbf {\bibinfo {volume}
  {81}},\ \bibinfo {pages} {042330} (\bibinfo {year} {2010})}\BibitemShut
  {NoStop}%
\bibitem [{\citenamefont {Childs}\ \emph {et~al.}(2013)\citenamefont {Childs},
  \citenamefont {Gosset},\ and\ \citenamefont {Webb}}]{childs13}%
  \BibitemOpen
  \bibfield  {author} {\bibinfo {author} {\bibfnamefont {A.~M.}\ \bibnamefont
  {Childs}}, \bibinfo {author} {\bibfnamefont {D.}~\bibnamefont {Gosset}},\
  and\ \bibinfo {author} {\bibfnamefont {Z.}~\bibnamefont {Webb}},\ }\bibfield
  {title} {\bibinfo {title} {Universal computation by multiparticle quantum
  walk},\ }\href {https://doi.org/10.1126/science.1229957} {\bibfield
  {journal} {\bibinfo  {journal} {Science}\ }\textbf {\bibinfo {volume}
  {339}},\ \bibinfo {pages} {791} (\bibinfo {year} {2013})}\BibitemShut
  {NoStop}%
\bibitem [{\citenamefont {S.}(2007)}]{bose07}%
  \BibitemOpen
  \bibfield  {author} {\bibinfo {author} {\bibfnamefont {B.}~\bibnamefont
  {S.}},\ }\bibfield  {title} {\bibinfo {title} {Quantum communication through
  spin chain dynamics: an introductory overview},\ }\href
  {https://doi.org/10.1080/00107510701342313} {\bibfield  {journal} {\bibinfo
  {journal} {Contemp. Phys.}\ }\textbf {\bibinfo {volume} {48}},\ \bibinfo
  {pages} {13} (\bibinfo {year} {2007})}\BibitemShut {NoStop}%
\bibitem [{\citenamefont {Farhi}\ and\ \citenamefont
  {Gutmann}(1998)}]{fahri98}%
  \BibitemOpen
  \bibfield  {author} {\bibinfo {author} {\bibfnamefont {E.}~\bibnamefont
  {Farhi}}\ and\ \bibinfo {author} {\bibfnamefont {S.}~\bibnamefont
  {Gutmann}},\ }\bibfield  {title} {\bibinfo {title} {Quantum computation and
  decision trees},\ }\href {https://doi.org/10.1103/PhysRevA.58.915} {\bibfield
   {journal} {\bibinfo  {journal} {Phys. Rev. A}\ }\textbf {\bibinfo {volume}
  {58}},\ \bibinfo {pages} {915} (\bibinfo {year} {1998})}\BibitemShut
  {NoStop}%
\bibitem [{\citenamefont {Kempe}(2003)}]{kempe003}%
  \BibitemOpen
  \bibfield  {author} {\bibinfo {author} {\bibfnamefont {J.}~\bibnamefont
  {Kempe}},\ }\bibfield  {title} {\bibinfo {title} {Quantum random walks: An
  introductory overview},\ }\href
  {https://doi.org/10.1080/00107151031000110776} {\bibfield  {journal}
  {\bibinfo  {journal} {Contemp. Phys.}\ }\textbf {\bibinfo {volume} {44}},\
  \bibinfo {pages} {307} (\bibinfo {year} {2003})}\BibitemShut {NoStop}%
\bibitem [{\citenamefont {Venegas-Andraca}(2012)}]{venegas2012}%
  \BibitemOpen
  \bibfield  {author} {\bibinfo {author} {\bibfnamefont {S.~E.}\ \bibnamefont
  {Venegas-Andraca}},\ }\bibfield  {title} {\bibinfo {title} {Quantum walks: a
  comprehensive review},\ }\href
  {https://doi.org/https://doi.org/10.1007/s11128-012-0432-5} {\bibfield
  {journal} {\bibinfo  {journal} {Quantum Inf. Process.}\ }\textbf {\bibinfo
  {volume} {11}},\ \bibinfo {pages} {1015} (\bibinfo {year}
  {2012})}\BibitemShut {NoStop}%
\bibitem [{\citenamefont {Kadian}\ \emph {et~al.}(2021)\citenamefont {Kadian},
  \citenamefont {Garhwal},\ and\ \citenamefont {Kumar}}]{kadian21}%
  \BibitemOpen
  \bibfield  {author} {\bibinfo {author} {\bibfnamefont {K.}~\bibnamefont
  {Kadian}}, \bibinfo {author} {\bibfnamefont {S.}~\bibnamefont {Garhwal}},\
  and\ \bibinfo {author} {\bibfnamefont {A.}~\bibnamefont {Kumar}},\ }\bibfield
   {title} {\bibinfo {title} {Quantum walk and its application domains: A
  systematic review},\ }\href
  {https://doi.org/https://doi.org/10.1016/j.cosrev.2021.100419} {\bibfield
  {journal} {\bibinfo  {journal} {Comput. Sci. Rev.}\ }\textbf {\bibinfo
  {volume} {41}},\ \bibinfo {pages} {100419} (\bibinfo {year}
  {2021})}\BibitemShut {NoStop}%
\bibitem [{\citenamefont {M{\"u}lken}\ \emph {et~al.}(2006)\citenamefont
  {M{\"u}lken}, \citenamefont {Bierbaum},\ and\ \citenamefont
  {Blumen}}]{Mulken:2006}%
  \BibitemOpen
  \bibfield  {author} {\bibinfo {author} {\bibfnamefont {O.}~\bibnamefont
  {M{\"u}lken}}, \bibinfo {author} {\bibfnamefont {V.}~\bibnamefont
  {Bierbaum}},\ and\ \bibinfo {author} {\bibfnamefont {A.}~\bibnamefont
  {Blumen}},\ }\bibfield  {title} {\bibinfo {title} {Coherent exciton transport
  in dendrimers and continuous-time quantum walks},\ }\href
  {https://doi.org/10.1063/1.2179427} {\bibfield  {journal} {\bibinfo
  {journal} {J. Chem. Phys.}\ }\textbf {\bibinfo {volume} {124}},\ \bibinfo
  {pages} {124905} (\bibinfo {year} {2006})}\BibitemShut {NoStop}%
\bibitem [{\citenamefont {A.}(2010)}]{kay10}%
  \BibitemOpen
  \bibfield  {author} {\bibinfo {author} {\bibfnamefont {K.}~\bibnamefont
  {A.}},\ }\bibfield  {title} {\bibinfo {title} {Perfect, efficient, state
  transfer and its application as a constructive tool},\ }\href
  {https://doi.org/10.1142/S0219749910006514} {\bibfield  {journal} {\bibinfo
  {journal} {Int. J. Quantum Inf.}\ }\textbf {\bibinfo {volume} {08}},\
  \bibinfo {pages} {641} (\bibinfo {year} {2010})}\BibitemShut {NoStop}%
\bibitem [{\citenamefont {Chudzicki}\ and\ \citenamefont
  {Strauch}(2010)}]{Chudzicki10}%
  \BibitemOpen
  \bibfield  {author} {\bibinfo {author} {\bibfnamefont {C.}~\bibnamefont
  {Chudzicki}}\ and\ \bibinfo {author} {\bibfnamefont {F.~W.}\ \bibnamefont
  {Strauch}},\ }\bibfield  {title} {\bibinfo {title} {Parallel state transfer
  and efficient quantum routing on quantum networks},\ }\href
  {https://doi.org/10.1103/PhysRevLett.105.260501} {\bibfield  {journal}
  {\bibinfo  {journal} {Phys. Rev. Lett.}\ }\textbf {\bibinfo {volume} {105}},\
  \bibinfo {pages} {260501} (\bibinfo {year} {2010})}\BibitemShut {NoStop}%
\bibitem [{\citenamefont {Kendon}\ and\ \citenamefont
  {Tamon}(2011)}]{kendon2011}%
  \BibitemOpen
  \bibfield  {author} {\bibinfo {author} {\bibfnamefont {V.~M.}\ \bibnamefont
  {Kendon}}\ and\ \bibinfo {author} {\bibfnamefont {C.}~\bibnamefont {Tamon}},\
  }\bibfield  {title} {\bibinfo {title} {Perfect state transfer in quantum
  walks on graphs},\ }\href
  {https://doi.org/https://doi.org/10.1166/jctn.2011.1706} {\bibfield
  {journal} {\bibinfo  {journal} {J. Comput. Theor. Nanos.}\ }\textbf {\bibinfo
  {volume} {8}},\ \bibinfo {pages} {422} (\bibinfo {year} {2011})}\BibitemShut
  {NoStop}%
\bibitem [{\citenamefont {Alvir}\ \emph {et~al.}(2016)\citenamefont {Alvir},
  \citenamefont {Dever}, \citenamefont {Lovitz}, \citenamefont {Myer},
  \citenamefont {Tamon}, \citenamefont {Xu},\ and\ \citenamefont
  {Zhan}}]{alvir2016}%
  \BibitemOpen
  \bibfield  {author} {\bibinfo {author} {\bibfnamefont {R.}~\bibnamefont
  {Alvir}}, \bibinfo {author} {\bibfnamefont {S.}~\bibnamefont {Dever}},
  \bibinfo {author} {\bibfnamefont {B.}~\bibnamefont {Lovitz}}, \bibinfo
  {author} {\bibfnamefont {J.}~\bibnamefont {Myer}}, \bibinfo {author}
  {\bibfnamefont {C.}~\bibnamefont {Tamon}}, \bibinfo {author} {\bibfnamefont
  {Y.}~\bibnamefont {Xu}},\ and\ \bibinfo {author} {\bibfnamefont
  {H.}~\bibnamefont {Zhan}},\ }\bibfield  {title} {\bibinfo {title} {Perfect
  state transfer in laplacian quantum walk},\ }\href
  {https://doi.org/https://doi.org/10.1007/s10801-015-0642-x} {\bibfield
  {journal} {\bibinfo  {journal} {J. Algebr. Comb.}\ }\textbf {\bibinfo
  {volume} {43}},\ \bibinfo {pages} {801} (\bibinfo {year} {2016})}\BibitemShut
  {NoStop}%
\bibitem [{\citenamefont {Cavazzoni}\ \emph {et~al.}(2022)\citenamefont
  {Cavazzoni}, \citenamefont {Razzoli}, \citenamefont {Bordone},\ and\
  \citenamefont {Paris}}]{cavazzoni22}%
  \BibitemOpen
  \bibfield  {author} {\bibinfo {author} {\bibfnamefont {S.}~\bibnamefont
  {Cavazzoni}}, \bibinfo {author} {\bibfnamefont {L.}~\bibnamefont {Razzoli}},
  \bibinfo {author} {\bibfnamefont {P.}~\bibnamefont {Bordone}},\ and\ \bibinfo
  {author} {\bibfnamefont {M.~G.~A.}\ \bibnamefont {Paris}},\ }\bibfield
  {title} {\bibinfo {title} {Perturbed graphs achieve unit transport efficiency
  without environmental noise},\ }\href
  {https://doi.org/10.1103/PhysRevE.106.024118} {\bibfield  {journal} {\bibinfo
   {journal} {Phys. Rev. E}\ }\textbf {\bibinfo {volume} {106}},\ \bibinfo
  {pages} {024118} (\bibinfo {year} {2022})}\BibitemShut {NoStop}%
\bibitem [{\citenamefont {Maquin{\'e}~Batalha}\ \emph
  {et~al.}(2022)\citenamefont {Maquin{\'e}~Batalha}, \citenamefont {Volta},
  \citenamefont {Strunz},\ and\ \citenamefont {Galiceanu}}]{Maquine22}%
  \BibitemOpen
  \bibfield  {author} {\bibinfo {author} {\bibfnamefont {G.}~\bibnamefont
  {Maquin{\'e}~Batalha}}, \bibinfo {author} {\bibfnamefont {A.}~\bibnamefont
  {Volta}}, \bibinfo {author} {\bibfnamefont {W.~T.}\ \bibnamefont {Strunz}},\
  and\ \bibinfo {author} {\bibfnamefont {M.}~\bibnamefont {Galiceanu}},\
  }\bibfield  {title} {\bibinfo {title} {Quantum transport on honeycomb
  networks},\ }\href {https://doi.org/10.1038/s41598-022-10537-w} {\bibfield
  {journal} {\bibinfo  {journal} {Sci. Rep.}\ }\textbf {\bibinfo {volume}
  {12}},\ \bibinfo {pages} {6896} (\bibinfo {year} {2022})}\BibitemShut
  {NoStop}%
\bibitem [{\citenamefont {Qiang}\ \emph {et~al.}(2016)\citenamefont {Qiang},
  \citenamefont {Loke}, \citenamefont {Montanaro}, \citenamefont
  {Aungskunsiri}, \citenamefont {Zhou}, \citenamefont {O'Brien}, \citenamefont
  {Wang},\ and\ \citenamefont {Matthews}}]{qiang16}%
  \BibitemOpen
  \bibfield  {author} {\bibinfo {author} {\bibfnamefont {X.}~\bibnamefont
  {Qiang}}, \bibinfo {author} {\bibfnamefont {T.}~\bibnamefont {Loke}},
  \bibinfo {author} {\bibfnamefont {A.}~\bibnamefont {Montanaro}}, \bibinfo
  {author} {\bibfnamefont {K.}~\bibnamefont {Aungskunsiri}}, \bibinfo {author}
  {\bibfnamefont {X.}~\bibnamefont {Zhou}}, \bibinfo {author} {\bibfnamefont
  {J.~L.}\ \bibnamefont {O'Brien}}, \bibinfo {author} {\bibfnamefont {J.~B.}\
  \bibnamefont {Wang}},\ and\ \bibinfo {author} {\bibfnamefont {J.~C.~F.}\
  \bibnamefont {Matthews}},\ }\bibfield  {title} {\bibinfo {title} {Efficient
  quantum walk on a quantum processor},\ }\href
  {https://doi.org/10.1038/ncomms11511} {\bibfield  {journal} {\bibinfo
  {journal} {Nat. Comm.}\ }\textbf {\bibinfo {volume} {7}},\ \bibinfo {pages}
  {11511} (\bibinfo {year} {2016})}\BibitemShut {NoStop}%
\bibitem [{\citenamefont {Tang}\ \emph {et~al.}(2018)\citenamefont {Tang},
  \citenamefont {Lin}, \citenamefont {Feng}, \citenamefont {Chen},
  \citenamefont {Gao}, \citenamefont {Sun}, \citenamefont {Wang}, \citenamefont
  {Lai}, \citenamefont {Xu}, \citenamefont {Wang}, \citenamefont {Qiao},
  \citenamefont {Yang},\ and\ \citenamefont {Jin}}]{tang18}%
  \BibitemOpen
  \bibfield  {author} {\bibinfo {author} {\bibfnamefont {H.}~\bibnamefont
  {Tang}}, \bibinfo {author} {\bibfnamefont {X.-F.}\ \bibnamefont {Lin}},
  \bibinfo {author} {\bibfnamefont {Z.}~\bibnamefont {Feng}}, \bibinfo {author}
  {\bibfnamefont {J.-Y.}\ \bibnamefont {Chen}}, \bibinfo {author}
  {\bibfnamefont {J.}~\bibnamefont {Gao}}, \bibinfo {author} {\bibfnamefont
  {K.}~\bibnamefont {Sun}}, \bibinfo {author} {\bibfnamefont {C.-Y.}\
  \bibnamefont {Wang}}, \bibinfo {author} {\bibfnamefont {P.-C.}\ \bibnamefont
  {Lai}}, \bibinfo {author} {\bibfnamefont {X.-Y.}\ \bibnamefont {Xu}},
  \bibinfo {author} {\bibfnamefont {Y.}~\bibnamefont {Wang}}, \bibinfo {author}
  {\bibfnamefont {L.-F.}\ \bibnamefont {Qiao}}, \bibinfo {author}
  {\bibfnamefont {A.-L.}\ \bibnamefont {Yang}},\ and\ \bibinfo {author}
  {\bibfnamefont {X.-M.}\ \bibnamefont {Jin}},\ }\bibfield  {title} {\bibinfo
  {title} {Experimental two-dimensional quantum walk on a photonic chip},\
  }\href {https://doi.org/10.1126/sciadv.aat3174} {\bibfield  {journal}
  {\bibinfo  {journal} {Science Adv.}\ }\textbf {\bibinfo {volume} {4}},\
  \bibinfo {pages} {eaat3174} (\bibinfo {year} {2018})}\BibitemShut {NoStop}%
\bibitem [{\citenamefont {Imany}\ \emph {et~al.}(2020)\citenamefont {Imany},
  \citenamefont {Lingaraju}, \citenamefont {Alshaykh}, \citenamefont {Leaird},\
  and\ \citenamefont {Weiner}}]{imany20}%
  \BibitemOpen
  \bibfield  {author} {\bibinfo {author} {\bibfnamefont {P.}~\bibnamefont
  {Imany}}, \bibinfo {author} {\bibfnamefont {N.~B.}\ \bibnamefont
  {Lingaraju}}, \bibinfo {author} {\bibfnamefont {M.~S.}\ \bibnamefont
  {Alshaykh}}, \bibinfo {author} {\bibfnamefont {D.~E.}\ \bibnamefont
  {Leaird}},\ and\ \bibinfo {author} {\bibfnamefont {A.~M.}\ \bibnamefont
  {Weiner}},\ }\bibfield  {title} {\bibinfo {title} {Probing quantum walks
  through coherent control of high-dimensionally entangled photons},\ }\href
  {https://doi.org/10.1126/sciadv.aba8066} {\bibfield  {journal} {\bibinfo
  {journal} {Science Adv.}\ }\textbf {\bibinfo {volume} {6}},\ \bibinfo {pages}
  {eaba8066} (\bibinfo {year} {2020})}\BibitemShut {NoStop}%
\bibitem [{\citenamefont {Meier}\ \emph {et~al.}(2016)\citenamefont {Meier},
  \citenamefont {An},\ and\ \citenamefont {Gadway}}]{meier16}%
  \BibitemOpen
  \bibfield  {author} {\bibinfo {author} {\bibfnamefont {E.~J.}\ \bibnamefont
  {Meier}}, \bibinfo {author} {\bibfnamefont {F.~A.}\ \bibnamefont {An}},\ and\
  \bibinfo {author} {\bibfnamefont {B.}~\bibnamefont {Gadway}},\ }\bibfield
  {title} {\bibinfo {title} {Atom-optics simulator of lattice transport
  phenomena},\ }\href {https://doi.org/10.1103/PhysRevA.93.051602} {\bibfield
  {journal} {\bibinfo  {journal} {Phys. Rev. A}\ }\textbf {\bibinfo {volume}
  {93}},\ \bibinfo {pages} {051602} (\bibinfo {year} {2016})}\BibitemShut
  {NoStop}%
\bibitem [{\citenamefont {Tamura}\ \emph {et~al.}(2020)\citenamefont {Tamura},
  \citenamefont {Mukaiyama},\ and\ \citenamefont {Toyoda}}]{tamura20}%
  \BibitemOpen
  \bibfield  {author} {\bibinfo {author} {\bibfnamefont {M.}~\bibnamefont
  {Tamura}}, \bibinfo {author} {\bibfnamefont {T.}~\bibnamefont {Mukaiyama}},\
  and\ \bibinfo {author} {\bibfnamefont {K.}~\bibnamefont {Toyoda}},\
  }\bibfield  {title} {\bibinfo {title} {Quantum walks of a phonon in trapped
  ions},\ }\bibfield  {journal} {\bibinfo  {journal} {Phys. Rev. Lett.}\
  }\textbf {\bibinfo {volume} {124}},\ \href
  {https://doi.org/10.1103/PhysRevLett.124.200501}
  {10.1103/PhysRevLett.124.200501} (\bibinfo {year} {2020})\BibitemShut
  {NoStop}%
\bibitem [{\citenamefont {Perets}\ \emph {et~al.}(2008)\citenamefont {Perets},
  \citenamefont {Lahini}, \citenamefont {Pozzi}, \citenamefont {Sorel},
  \citenamefont {Morandotti},\ and\ \citenamefont {Silberberg}}]{perets08}%
  \BibitemOpen
  \bibfield  {author} {\bibinfo {author} {\bibfnamefont {H.~B.}\ \bibnamefont
  {Perets}}, \bibinfo {author} {\bibfnamefont {Y.}~\bibnamefont {Lahini}},
  \bibinfo {author} {\bibfnamefont {F.}~\bibnamefont {Pozzi}}, \bibinfo
  {author} {\bibfnamefont {M.}~\bibnamefont {Sorel}}, \bibinfo {author}
  {\bibfnamefont {R.}~\bibnamefont {Morandotti}},\ and\ \bibinfo {author}
  {\bibfnamefont {Y.}~\bibnamefont {Silberberg}},\ }\bibfield  {title}
  {\bibinfo {title} {Realization of quantum walks with negligible decoherence
  in waveguide lattices},\ }\href
  {https://doi.org/10.1103/PhysRevLett.100.170506} {\bibfield  {journal}
  {\bibinfo  {journal} {Phys. Rev. Lett.}\ }\textbf {\bibinfo {volume} {100}},\
  \bibinfo {pages} {170506} (\bibinfo {year} {2008})}\BibitemShut {NoStop}%
\bibitem [{\citenamefont {Poulios}\ \emph {et~al.}(2014)\citenamefont
  {Poulios}, \citenamefont {Keil}, \citenamefont {Fry}, \citenamefont
  {Meinecke}, \citenamefont {Matthews}, \citenamefont {Politi}, \citenamefont
  {Lobino}, \citenamefont {Gr\"afe}, \citenamefont {Heinrich}, \citenamefont
  {Nolte}, \citenamefont {Szameit},\ and\ \citenamefont {O'Brien}}]{poulios14}%
  \BibitemOpen
  \bibfield  {author} {\bibinfo {author} {\bibfnamefont {K.}~\bibnamefont
  {Poulios}}, \bibinfo {author} {\bibfnamefont {R.}~\bibnamefont {Keil}},
  \bibinfo {author} {\bibfnamefont {D.}~\bibnamefont {Fry}}, \bibinfo {author}
  {\bibfnamefont {J.~D.~A.}\ \bibnamefont {Meinecke}}, \bibinfo {author}
  {\bibfnamefont {J.~C.~F.}\ \bibnamefont {Matthews}}, \bibinfo {author}
  {\bibfnamefont {A.}~\bibnamefont {Politi}}, \bibinfo {author} {\bibfnamefont
  {M.}~\bibnamefont {Lobino}}, \bibinfo {author} {\bibfnamefont
  {M.}~\bibnamefont {Gr\"afe}}, \bibinfo {author} {\bibfnamefont
  {M.}~\bibnamefont {Heinrich}}, \bibinfo {author} {\bibfnamefont
  {S.}~\bibnamefont {Nolte}}, \bibinfo {author} {\bibfnamefont
  {A.}~\bibnamefont {Szameit}},\ and\ \bibinfo {author} {\bibfnamefont {J.~L.}\
  \bibnamefont {O'Brien}},\ }\bibfield  {title} {\bibinfo {title} {Quantum
  walks of correlated photon pairs in two-dimensional waveguide arrays},\
  }\href {https://doi.org/10.1103/PhysRevLett.112.143604} {\bibfield  {journal}
  {\bibinfo  {journal} {Phys. Rev. Lett.}\ }\textbf {\bibinfo {volume} {112}},\
  \bibinfo {pages} {143604} (\bibinfo {year} {2014})}\BibitemShut {NoStop}%
\bibitem [{\citenamefont {Caruso}\ \emph {et~al.}(2016)\citenamefont {Caruso},
  \citenamefont {Crespi}, \citenamefont {Ciriolo}, \citenamefont {Sciarrino},\
  and\ \citenamefont {Osellame}}]{caruso16}%
  \BibitemOpen
  \bibfield  {author} {\bibinfo {author} {\bibfnamefont {F.}~\bibnamefont
  {Caruso}}, \bibinfo {author} {\bibfnamefont {A.}~\bibnamefont {Crespi}},
  \bibinfo {author} {\bibfnamefont {A.~G.}\ \bibnamefont {Ciriolo}}, \bibinfo
  {author} {\bibfnamefont {F.}~\bibnamefont {Sciarrino}},\ and\ \bibinfo
  {author} {\bibfnamefont {R.}~\bibnamefont {Osellame}},\ }\bibfield  {title}
  {\bibinfo {title} {Fast escape of a quantum walker from an integrated
  photonic maze},\ }\href {https://doi.org/10.1038/ncomms11682} {\bibfield
  {journal} {\bibinfo  {journal} {Nat. Commun.}\ }\textbf {\bibinfo {volume}
  {7}},\ \bibinfo {pages} {11682} (\bibinfo {year} {2016})}\BibitemShut
  {NoStop}%
\bibitem [{\citenamefont {Benedetti}\ \emph {et~al.}(2021)\citenamefont
  {Benedetti}, \citenamefont {Tamascelli}, \citenamefont {Paris},\ and\
  \citenamefont {Crespi}}]{benedetti21}%
  \BibitemOpen
  \bibfield  {author} {\bibinfo {author} {\bibfnamefont {C.}~\bibnamefont
  {Benedetti}}, \bibinfo {author} {\bibfnamefont {D.}~\bibnamefont
  {Tamascelli}}, \bibinfo {author} {\bibfnamefont {M.~G.}\ \bibnamefont
  {Paris}},\ and\ \bibinfo {author} {\bibfnamefont {A.}~\bibnamefont
  {Crespi}},\ }\bibfield  {title} {\bibinfo {title} {Quantum spatial search in
  two-dimensional waveguide arrays},\ }\href
  {https://doi.org/10.1103/PhysRevApplied.16.054036} {\bibfield  {journal}
  {\bibinfo  {journal} {Phys. Rev. Applied}\ }\textbf {\bibinfo {volume}
  {16}},\ \bibinfo {pages} {054036} (\bibinfo {year} {2021})}\BibitemShut
  {NoStop}%
\bibitem [{\citenamefont {B\"ohm}\ \emph {et~al.}(2015)\citenamefont {B\"ohm},
  \citenamefont {Bellec}, \citenamefont {Mortessagne}, \citenamefont {Kuhl},
  \citenamefont {Barkhofen}, \citenamefont {Gehler}, \citenamefont
  {St\"ockmann}, \citenamefont {Foulger}, \citenamefont {Gnutzmann},\ and\
  \citenamefont {Tanner}}]{bohm15}%
  \BibitemOpen
  \bibfield  {author} {\bibinfo {author} {\bibfnamefont {J.}~\bibnamefont
  {B\"ohm}}, \bibinfo {author} {\bibfnamefont {M.}~\bibnamefont {Bellec}},
  \bibinfo {author} {\bibfnamefont {F.}~\bibnamefont {Mortessagne}}, \bibinfo
  {author} {\bibfnamefont {U.}~\bibnamefont {Kuhl}}, \bibinfo {author}
  {\bibfnamefont {S.}~\bibnamefont {Barkhofen}}, \bibinfo {author}
  {\bibfnamefont {S.}~\bibnamefont {Gehler}}, \bibinfo {author} {\bibfnamefont
  {H.-J.}\ \bibnamefont {St\"ockmann}}, \bibinfo {author} {\bibfnamefont
  {I.}~\bibnamefont {Foulger}}, \bibinfo {author} {\bibfnamefont
  {S.}~\bibnamefont {Gnutzmann}},\ and\ \bibinfo {author} {\bibfnamefont
  {G.}~\bibnamefont {Tanner}},\ }\bibfield  {title} {\bibinfo {title}
  {Microwave experiments simulating quantum search and directed transport in
  artificial graphene},\ }\href
  {https://doi.org/10.1103/PhysRevLett.114.110501} {\bibfield  {journal}
  {\bibinfo  {journal} {Phys. Rev. Lett.}\ }\textbf {\bibinfo {volume} {114}},\
  \bibinfo {pages} {110501} (\bibinfo {year} {2015})}\BibitemShut {NoStop}%
\bibitem [{\citenamefont {Du}\ \emph {et~al.}(2003)\citenamefont {Du},
  \citenamefont {Li}, \citenamefont {Xu}, \citenamefont {Shi}, \citenamefont
  {Wu}, \citenamefont {Zhou},\ and\ \citenamefont {Han}}]{du03}%
  \BibitemOpen
  \bibfield  {author} {\bibinfo {author} {\bibfnamefont {J.}~\bibnamefont
  {Du}}, \bibinfo {author} {\bibfnamefont {H.}~\bibnamefont {Li}}, \bibinfo
  {author} {\bibfnamefont {X.}~\bibnamefont {Xu}}, \bibinfo {author}
  {\bibfnamefont {M.}~\bibnamefont {Shi}}, \bibinfo {author} {\bibfnamefont
  {J.}~\bibnamefont {Wu}}, \bibinfo {author} {\bibfnamefont {X.}~\bibnamefont
  {Zhou}},\ and\ \bibinfo {author} {\bibfnamefont {R.}~\bibnamefont {Han}},\
  }\bibfield  {title} {\bibinfo {title} {Experimental implementation of the
  quantum random-walk algorithm},\ }\href
  {https://doi.org/10.1103/PhysRevA.67.042316} {\bibfield  {journal} {\bibinfo
  {journal} {Phys. Rev. A}\ }\textbf {\bibinfo {volume} {67}},\ \bibinfo
  {pages} {042316} (\bibinfo {year} {2003})}\BibitemShut {NoStop}%
\bibitem [{\citenamefont {Zimbor{\'a}s}\ \emph {et~al.}(2013)\citenamefont
  {Zimbor{\'a}s}, \citenamefont {Faccin}, \citenamefont {K{\'a}d{\'a}r},
  \citenamefont {Whitfield}, \citenamefont {Lanyon},\ and\ \citenamefont
  {Biamonte}}]{Zimboras13}%
  \BibitemOpen
  \bibfield  {author} {\bibinfo {author} {\bibfnamefont {Z.}~\bibnamefont
  {Zimbor{\'a}s}}, \bibinfo {author} {\bibfnamefont {M.}~\bibnamefont
  {Faccin}}, \bibinfo {author} {\bibfnamefont {Z.}~\bibnamefont
  {K{\'a}d{\'a}r}}, \bibinfo {author} {\bibfnamefont {J.~D.}\ \bibnamefont
  {Whitfield}}, \bibinfo {author} {\bibfnamefont {B.~P.}\ \bibnamefont
  {Lanyon}},\ and\ \bibinfo {author} {\bibfnamefont {J.}~\bibnamefont
  {Biamonte}},\ }\bibfield  {title} {\bibinfo {title} {Quantum transport
  enhancement by time-reversal symmetry breaking},\ }\href
  {https://doi.org/10.1038/srep02361} {\bibfield  {journal} {\bibinfo
  {journal} {Scientific Reports}\ }\textbf {\bibinfo {volume} {3}},\ \bibinfo
  {pages} {2361} (\bibinfo {year} {2013})}\BibitemShut {NoStop}%
\bibitem [{\citenamefont {Lu}\ \emph {et~al.}(2016)\citenamefont {Lu},
  \citenamefont {Biamonte}, \citenamefont {Li}, \citenamefont {Li},
  \citenamefont {Johnson}, \citenamefont {Bergholm}, \citenamefont {Faccin},
  \citenamefont {Zimbor\'as}, \citenamefont {Laflamme}, \citenamefont {Baugh},\
  and\ \citenamefont {Lloyd}}]{lu16}%
  \BibitemOpen
  \bibfield  {author} {\bibinfo {author} {\bibfnamefont {D.}~\bibnamefont
  {Lu}}, \bibinfo {author} {\bibfnamefont {J.~D.}\ \bibnamefont {Biamonte}},
  \bibinfo {author} {\bibfnamefont {J.}~\bibnamefont {Li}}, \bibinfo {author}
  {\bibfnamefont {H.}~\bibnamefont {Li}}, \bibinfo {author} {\bibfnamefont
  {T.~H.}\ \bibnamefont {Johnson}}, \bibinfo {author} {\bibfnamefont
  {V.}~\bibnamefont {Bergholm}}, \bibinfo {author} {\bibfnamefont
  {M.}~\bibnamefont {Faccin}}, \bibinfo {author} {\bibfnamefont
  {Z.}~\bibnamefont {Zimbor\'as}}, \bibinfo {author} {\bibfnamefont
  {R.}~\bibnamefont {Laflamme}}, \bibinfo {author} {\bibfnamefont
  {J.}~\bibnamefont {Baugh}},\ and\ \bibinfo {author} {\bibfnamefont
  {S.}~\bibnamefont {Lloyd}},\ }\bibfield  {title} {\bibinfo {title} {Chiral
  quantum walks},\ }\href {https://doi.org/10.1103/PhysRevA.93.042302}
  {\bibfield  {journal} {\bibinfo  {journal} {Phys. Rev. A}\ }\textbf {\bibinfo
  {volume} {93}},\ \bibinfo {pages} {042302} (\bibinfo {year}
  {2016})}\BibitemShut {NoStop}%
\bibitem [{\citenamefont {Frigerio}\ \emph {et~al.}(2021)\citenamefont
  {Frigerio}, \citenamefont {Benedetti}, \citenamefont {Olivares},\ and\
  \citenamefont {Paris}}]{frigerio21}%
  \BibitemOpen
  \bibfield  {author} {\bibinfo {author} {\bibfnamefont {M.}~\bibnamefont
  {Frigerio}}, \bibinfo {author} {\bibfnamefont {C.}~\bibnamefont {Benedetti}},
  \bibinfo {author} {\bibfnamefont {S.}~\bibnamefont {Olivares}},\ and\
  \bibinfo {author} {\bibfnamefont {M.~G.~A.}\ \bibnamefont {Paris}},\
  }\bibfield  {title} {\bibinfo {title} {Generalized quantum-classical
  correspondence for random walks on graphs},\ }\href
  {https://doi.org/10.1103/PhysRevA.104.L030201} {\bibfield  {journal}
  {\bibinfo  {journal} {Phys. Rev. A}\ }\textbf {\bibinfo {volume} {104}},\
  \bibinfo {pages} {L030201} (\bibinfo {year} {2021})}\BibitemShut {NoStop}%
\bibitem [{\citenamefont {Khalique}\ \emph {et~al.}(2021)\citenamefont
  {Khalique}, \citenamefont {A.}, \citenamefont {Wang},\ and\ \citenamefont
  {Twamley}}]{Khalique21}%
  \BibitemOpen
  \bibfield  {author} {\bibinfo {author} {\bibfnamefont {A.}~\bibnamefont
  {Khalique}}, \bibinfo {author} {\bibfnamefont {S.}~\bibnamefont {A.}},
  \bibinfo {author} {\bibfnamefont {J.}~\bibnamefont {Wang}},\ and\ \bibinfo
  {author} {\bibfnamefont {J.}~\bibnamefont {Twamley}},\ }\bibfield  {title}
  {\bibinfo {title} {Controlled information transfer in continuous-time chiral
  quantum walks},\ }\href {https://doi.org/10.1088/1367-2630/ac1551} {\bibfield
   {journal} {\bibinfo  {journal} {New J. Phys.}\ }\textbf {\bibinfo {volume}
  {23}},\ \bibinfo {pages} {083005} (\bibinfo {year} {2021})}\BibitemShut
  {NoStop}%
\bibitem [{\citenamefont {Frigerio}(2022)}]{frigerio22}%
  \BibitemOpen
  \bibfield  {author} {\bibinfo {author} {\bibfnamefont {M.~G.~A.}\
  \bibnamefont {Frigerio}, \bibfnamefont {M.~Paris}},\ }\bibfield  {title}
  {\bibinfo {title} {Swift chiral quantum walks},\ }\bibfield  {journal}
  {\bibinfo  {journal} {arXiv.2207.05168}\ }\href
  {https://doi.org/10.48550/arXiv.2207.05168} {10.48550/arXiv.2207.05168}
  (\bibinfo {year} {2022})\BibitemShut {NoStop}%
\bibitem [{\citenamefont {Yang}\ \emph {et~al.}(2015)\citenamefont {Yang},
  \citenamefont {Pan}, \citenamefont {Sun},\ and\ \citenamefont {Xu}}]{yang15}%
  \BibitemOpen
  \bibfield  {author} {\bibinfo {author} {\bibfnamefont {Y.-G.}\ \bibnamefont
  {Yang}}, \bibinfo {author} {\bibfnamefont {Q.-X.}\ \bibnamefont {Pan}},
  \bibinfo {author} {\bibfnamefont {S.-J.}\ \bibnamefont {Sun}},\ and\ \bibinfo
  {author} {\bibfnamefont {P.}~\bibnamefont {Xu}},\ }\bibfield  {title}
  {\bibinfo {title} {Novel image encryption based on quantum walks},\ }\href
  {https://doi.org/10.1038/srep07784} {\bibfield  {journal} {\bibinfo
  {journal} {Sci. Rep.}\ }\textbf {\bibinfo {volume} {5}},\ \bibinfo {pages}
  {7784} (\bibinfo {year} {2015})}\BibitemShut {NoStop}%
\bibitem [{\citenamefont {{Abd EL-Latif}}\ \emph {et~al.}(2019)\citenamefont
  {{Abd EL-Latif}}, \citenamefont {Abd-El-Atty},\ and\ \citenamefont
  {Venegas-Andraca}}]{abdellatif2019}%
  \BibitemOpen
  \bibfield  {author} {\bibinfo {author} {\bibfnamefont {A.~A.}\ \bibnamefont
  {{Abd EL-Latif}}}, \bibinfo {author} {\bibfnamefont {B.}~\bibnamefont
  {Abd-El-Atty}},\ and\ \bibinfo {author} {\bibfnamefont {S.~E.}\ \bibnamefont
  {Venegas-Andraca}},\ }\bibfield  {title} {\bibinfo {title} {A novel image
  steganography technique based on quantum substitution boxes},\ }\href
  {https://doi.org/https://doi.org/10.1016/j.optlastec.2019.03.005} {\bibfield
  {journal} {\bibinfo  {journal} {Opt. Laser Technol.}\ }\textbf {\bibinfo
  {volume} {116}},\ \bibinfo {pages} {92} (\bibinfo {year} {2019})}\BibitemShut
  {NoStop}%
\bibitem [{\citenamefont {A.}\ \emph {et~al.}(2020)\citenamefont {A.},
  \citenamefont {Abd-El-Atty},\ and\ \citenamefont
  {Venegas-Andraca}}]{ellatif2020}%
  \BibitemOpen
  \bibfield  {author} {\bibinfo {author} {\bibfnamefont {A.~E.-L.~A.}\
  \bibnamefont {A.}}, \bibinfo {author} {\bibfnamefont {B.}~\bibnamefont
  {Abd-El-Atty}},\ and\ \bibinfo {author} {\bibfnamefont {S.~E.}\ \bibnamefont
  {Venegas-Andraca}},\ }\bibfield  {title} {\bibinfo {title} {Controlled
  alternate quantum walk-based pseudo-random number generator and its
  application to quantum color image encryption},\ }\href
  {https://doi.org/https://doi.org/10.1016/j.physa.2019.123869} {\bibfield
  {journal} {\bibinfo  {journal} {Phys. A: Stat. Mech. Appl.}\ }\textbf
  {\bibinfo {volume} {547}},\ \bibinfo {pages} {123869} (\bibinfo {year}
  {2020})}\BibitemShut {NoStop}%
\bibitem [{\citenamefont {Tamascelli}\ \emph {et~al.}(2016)\citenamefont
  {Tamascelli}, \citenamefont {Benedetti}, \citenamefont {Olivares},\ and\
  \citenamefont {Paris}}]{tamascelli16}%
  \BibitemOpen
  \bibfield  {author} {\bibinfo {author} {\bibfnamefont {D.}~\bibnamefont
  {Tamascelli}}, \bibinfo {author} {\bibfnamefont {C.}~\bibnamefont
  {Benedetti}}, \bibinfo {author} {\bibfnamefont {S.}~\bibnamefont
  {Olivares}},\ and\ \bibinfo {author} {\bibfnamefont {M.~G.~A.}\ \bibnamefont
  {Paris}},\ }\bibfield  {title} {\bibinfo {title} {Characterization of qubit
  chains by feynman probes},\ }\href
  {https://doi.org/10.1103/PhysRevA.94.042129} {\bibfield  {journal} {\bibinfo
  {journal} {Phys. Rev. A}\ }\textbf {\bibinfo {volume} {94}},\ \bibinfo
  {pages} {042129} (\bibinfo {year} {2016})}\BibitemShut {NoStop}%
\bibitem [{\citenamefont {Seveso}\ \emph {et~al.}(2019)\citenamefont {Seveso},
  \citenamefont {Benedetti},\ and\ \citenamefont {Paris}}]{Seveso_2019}%
  \BibitemOpen
  \bibfield  {author} {\bibinfo {author} {\bibfnamefont {L.}~\bibnamefont
  {Seveso}}, \bibinfo {author} {\bibfnamefont {C.}~\bibnamefont {Benedetti}},\
  and\ \bibinfo {author} {\bibfnamefont {M.~G.~A.}\ \bibnamefont {Paris}},\
  }\bibfield  {title} {\bibinfo {title} {The walker speaks its graph: global
  and nearly-local probing of the tunnelling amplitude in continuous-time
  quantum walks},\ }\href {https://doi.org/10.1088/1751-8121/ab0195} {\bibfield
   {journal} {\bibinfo  {journal} {J. Phys. A: Math. Theo.}\ }\textbf {\bibinfo
  {volume} {52}},\ \bibinfo {pages} {105304} (\bibinfo {year}
  {2019})}\BibitemShut {NoStop}%
\bibitem [{\citenamefont {Gianani}\ \emph {et~al.}(2020)\citenamefont
  {Gianani}, \citenamefont {Genoni},\ and\ \citenamefont
  {Barbieri}}]{gianani20}%
  \BibitemOpen
  \bibfield  {author} {\bibinfo {author} {\bibfnamefont {I.}~\bibnamefont
  {Gianani}}, \bibinfo {author} {\bibfnamefont {M.~G.}\ \bibnamefont
  {Genoni}},\ and\ \bibinfo {author} {\bibfnamefont {M.}~\bibnamefont
  {Barbieri}},\ }\bibfield  {title} {\bibinfo {title} {Assessing data
  postprocessing for quantum estimation},\ }\href
  {https://doi.org/10.1109/JSTQE.2020.2982976} {\bibfield  {journal} {\bibinfo
  {journal} {IEEE Journal of Selected Topics in Quantum Electronics}\ }\textbf
  {\bibinfo {volume} {26}},\ \bibinfo {pages} {1} (\bibinfo {year}
  {2020})}\BibitemShut {NoStop}%
\bibitem [{\citenamefont {Spall}(2003)}]{spall03}%
  \BibitemOpen
  \bibfield  {author} {\bibinfo {author} {\bibfnamefont {J.}~\bibnamefont
  {Spall}},\ }\bibfield  {title} {\bibinfo {title} {Estimation via markov chain
  monte carlo},\ }\href {https://doi.org/10.1109/MCS.2003.1188770} {\bibfield
  {journal} {\bibinfo  {journal} {IEEE Control Systems Magazine}\ }\textbf
  {\bibinfo {volume} {23}},\ \bibinfo {pages} {34} (\bibinfo {year}
  {2003})}\BibitemShut {NoStop}%
\bibitem [{\citenamefont {Jacob}\ \emph {et~al.}(2020)\citenamefont {Jacob},
  \citenamefont {O'Leary},\ and\ \citenamefont {Atchad{\'{e}}}}]{Jacob2020}%
  \BibitemOpen
  \bibfield  {author} {\bibinfo {author} {\bibfnamefont {P.~E.}\ \bibnamefont
  {Jacob}}, \bibinfo {author} {\bibfnamefont {J.}~\bibnamefont {O'Leary}},\
  and\ \bibinfo {author} {\bibfnamefont {Y.~F.}\ \bibnamefont
  {Atchad{\'{e}}}},\ }\bibfield  {title} {\bibinfo {title} {{Unbiased Markov
  chain Monte Carlo methods with couplings}},\ }\href
  {https://doi.org/https://doi.org/10.1111/rssb.12336} {\bibfield  {journal}
  {\bibinfo  {journal} {J. R. Stat. Soc., B: Stat. Methodol.}\ }\textbf
  {\bibinfo {volume} {82}},\ \bibinfo {pages} {543} (\bibinfo {year}
  {2020})}\BibitemShut {NoStop}%
\end{thebibliography}%



\end{document}